**From Top to Bottom, Simple and Cost-Effective Methods to Nanopattern and Manufacture Anti-Delaminating, Thermally Stable Platforms on Kapton HN Flexible Films Using Inkjet Printing Technology to Produce Printable Nitrate Sensors, Mercury Aptasensors, Protein Sensors, and p-Type Organic Thin Film Transistors**

*Li-Kai Lin,** *Jung-Ting Tsai,*† *Susana Díaz-Amaya,*† *Muhammed R. Oduncu,*† *Yifan Zhang,*† *Peng-Yuan Huang,*† *Carlos Ostos,*† *Jacob P. Schmelzel,*† *Raheleh Mohammadrahimi,*† *Pengyu Xu, Nithin Raghunathan, Xinghang Zhang, Alexander Wei, David Bahr, Dimitrios Peroulis, and Lia A. Stanciu**

Dr. L. -K. Lin, M. R. Oduncu, Dr. S. Díaz-Amaya, Prof. L. A. Stanciu
Department of Materials Engineering, Purdue University. Neil Armstrong Hall of Engineering. 701 West Stadium Avenue, West Lafayette, IN, 47907-2045, USA
Birck Nanotechnology Center, Purdue University. 1205 W State St, West Lafayette, IN, 47907,USA
E-mails: lin688@purdue.edu, lstanciu@purdue.edu

J. -T. Tsai, Dr. R. Mohammadrahimi, Dr. Y. Zhang, P. Xu, Prof. X. Zhang, Prof. D. Bahr
Department of Materials Engineering, Purdue University. Neil Armstrong Hall of Engineering. 701 West Stadium Avenue, West Lafayette, IN, 47907-2045, USA

P. -Y. Huang
School of Library and Information Studies, University of Wisconsin–Madison, Helen C. White Hall, 600 N. Park Street, Madison, WI, 53706 USA

Prof. C. Ostos
CATALAD Research Group, Instituto de Química, Facultad de Ciencias Exactas y Naturales, Universidad de Antioquia UdeA Calle 70 No. 52-21, Medellín, Colombia

J. P. Schmelzel
School of Electrical and Computer Engineering Electrical Engineering Building, 465 Northwestern Ave. West Lafayette, IN 47907-2035, USA

Dr. N. Raghunathan
Birck Nanotechnology Center, Purdue University. 1205 W State St, West Lafayette, IN, 47907,USA

Prof. A. Wei
Department of Materials Engineering, Purdue University. Neil Armstrong Hall of Engineering. 701 West Stadium Avenue, West Lafayette, IN, 47907-2045, USA
Birck Nanotechnology Center, Purdue University. 1205 W State St, West Lafayette, IN, 47907,USA
Department of Chemistry, Purdue University, West Lafayette, 560 Oval Drive, Indiana, USA

Prof. D. Peroulis
School of Electrical and Computer Engineering Electrical Engineering Building, 465 Northwestern Ave. West Lafayette, IN 47907-2035, USA
Birck Nanotechnology Center, Purdue University. 1205 W State St, West Lafayette, IN, 47907,USA

[†]These authors contributed equally as the second author

Abstract

Kapton HN films, adopted worldwide due to their superior thermal durability (up to 400 °C), allows the high-temperature sintering of nanoparticle-based metal inks. By carefully selecting inks and Kapton substrates, outstanding thermal stability and anti-delaminating features are obtained in both aqueous and organic solutions and were applied to four novel devices: a solid-state ion-selective nitrate sensor, an ssDNA-based mercury aptasensor, a low-cost protein sensor, and a long-lasting organic thin film transistor (OTFT). Many experimental studies on parameter combinations were conducted during the development of the above devices. The results showed that the ion-selective nitrate sensor displayed a linear sensitivity range between $10^{-4.5}$ M and $10^{-1}$ M with a limit of detection of 2 ppm $NO_3^-$. The mercury sensor exhibited a linear correlation between the $R_{CT}$ values and the increasing concentrations of $Hg^{2+}$. The protein printed circuit board (PCB) sensor provided a much simpler method of protein detection. Finally, the OTFT demonstrated a stable performance with mobility values for the linear and saturation regimes being 0.0083 $cm^2V^{-1}S^{-1}$ and 0.0237 $cm^2V^{-1}S^{-1}$, respectively, and the threshold voltage being -6.75 V. These devices have shown their value and reveal possibilities that could be pursued.

## 1. Introduction

The field of printed electronics has grown spectacularly because of its promise for low-cost and high-performance devices that can be used in a broad range of applications. Printing technologies, often utilized to fabricate electronics in a simple and economical way, are currently popular because time and budget concerns have drastically increased in the era of technology. Among all printing methods, inkjet printing has a more simplified manufacturing procedure with the benefits of no contact writing and no troublesome coating and masking steps. These benefits make inkjet printing better fit the needs of new technologies that require a cost-effective and easy method for making a variety of applications. Printed electrochemical platforms are especially appealing, as they can be developed for in-field testing or as single-use sensors for point-of-care applications.[1–3]

Currently, various conductive inks are common and popular in the market. Among the numerous choices, silver and gold continually draw our attention. Silver is one of the most common metals that is frequently used to fabricate integrated circuits and biosensors because of its high conductivity.[4,5] On the other hand, gold plays a significant and irreplaceable role in biosensor fabrication owing to the variety of applications related to strong gold-thiol bonds, which are commonly used for the fixation of biomaterials. However, many of the discovered applications were fabricated by a troublesome and time-consuming photolithography process.[6–11]

Concerning the posttreatment of flexible inkjet-printed circuits or electrodes for attaining conductivity, the sintering/curing process is always a serious issue and is frequently limited by the applied substrates; thus, the application of printed electronics is restricted and determined by the thermal stability of the flexible substrates. A high temperature is constantly required by most nanoparticle-based inks. However, a high temperature is not compatible with most flexible substrates, such as polyethylene terephthalate (PET) and polyethylene naphthalate (PEN), with both having glass transition temperatures that are below 150 °C. In addition, although a polyimide-based substrate, such as a Kapton film, has high thermal resistance, each type of Kapton film has a different optimal temperature range depending on different application requirements. Therefore, carefully selected flexible substrates can broaden the range of applications available to inkjet-printed electronics.[4]

Over the past decades, mercury ($Hg^{2+}$) has been identified as a chemical threat to our environment and health, as well as nitrate ions. Regarding toxic mercury ions, tons of industrial waste that come from numerous human activities result in severe environmental crises worldwide, which slowly change the ecosystem. To stop this pollution from getting worse, a cost-effective sensing system for mercury is urgently needed. On the other hand, as one of several molecule-level contaminants, the nitrate ion ($NO^{3-}$) exists extensively in the environment because of its extreme solubility. It is important to control $NO_3^-$ levels, as excessive nitrate in agricultural runoff can pollute groundwater.[12–15] In addition, protein detection always plays a crucial role in most biomedical fields. However, most of the assays are fairly complicated, and the sensing materials are incredibly expensive. Therefore, a simple method for protein detection is highly desired.

In the field of IC fabrication, organic electronics have been increasingly studied since the 1960s, and the manufacture of organic thin film transistors has been a hot topic since a transistor is the most important component in integrated circuits.[16–22]

In this study, four different applications, including a solid-state nitrate sensor, a mercury aptasensor, a portable protein sensor, and an organic thin film transistor, which are commonly used in the fields of both bioscience and IC fabrication, are introduced. Based on the importance of the topics described, we carefully selected the best fit materials (25% Au inks for a low price, 40% Ag inks for high conductivity and good integrity, and a Kapton NH film for good flexibility and thermal stability) and prudently designed the manufacturing process to substantially decrease the cost (eliminating the need for laser sintering and photolithography) and increase the convenience of manufacturing microelectronics (**Figure 1**).[5]

## 2. Results and Discussion

### 2.1. Inkjet Printing Inks and Printing Parameter Optimization

Various nanoparticle inks have been widely used in different fields. Because nanoparticles have a high surface energy, organic surfactants are necessary to stabilize the surfaces. The presence of organics provides the printing pattern a high sintering temperature. In this study, two kinds of nanoparticle-based printing inks were studied to fabricate stable sensors in solutions and long-lasting organic thin film transistors. In **Figure 2A**, the silver ink composed of silver nanoflakes and the gold ink composed of gold nanoparticles were examined by transmission electron microscopy (TEM). The 50 nm silver agglomerates were composed of approximately 15 nm silver nanoflakes, which are shown in Figure 2A(a-c), while 100 nm gold agglomerates that were composed of approximately 5 nm gold nanoparticles are shown in Figure 2A(d-f).

The printing trace on the Kapton substrate had to be optimized before practically fabricating the sensor platforms and OTFT. Thus, the printing parameters, namely, the printing resolution, printing direction, saber angle, and drop spacing, were optimized. Figure 2B shows the optimization of the printing resolution for the silver and gold ink, with printing resolutions of 1693, 1270, 1016, 725, 564, 461, 390, and 338 dpi and digital printing patterns with widths of 20, 40, 80, 160, 240, 320, 400, 480, and 560 μm. The resulting printing traces are shown in Figure 2B (upper) for the silver traces and Figure 2B (lower) for the gold traces under a 10x optical microscope. Broken printed lines were observed in both the silver and gold traces when the printing resolution was below 564 dpi, while bulging phenomena were observed in both the silver and gold traces above 1016 dpi. Therefore, it is believed that 725 dpi is the optimal printing resolution for both inks on Kapton HN films, and this resolution was adopted for the subsequent fabrication processes.

### 2.2. Inkjet-Printed Film Adhesion Test for the Silver/Gold Patterns in Aqueous and Organic Solvents

To confirm the stability and integrity of the printed patterns before being practically used, the resulting Ag/Au patterns were tested under drastic conditions. Four groups of square Ag/Au patterns (0.73 cm sides) were printed on Kapton substrates. All Ag and Au patterns were then sonicated for 10 min and immersed in deionized water (**Figure 3A and C**) and ethanol (Figure 3B and D) for 12 days. The integrity and resistance of the immersed electrodes were observed and measured each day for 12 days. The resistance measurements were obtained with the two-probe method at diagonal vertices (Figure 3J-M), in which the resistance for these groups was stable within a week and most of them maintained stable resistance for 12 days, while only one of the groups had a slightly increasing resistance. The stable electrical property performances of the prepared films demonstrated sufficient film integrity for most biochemistry experiments in basic aqueous and organic solutions. The film integrity was examined by a standard adhesion test. In the test, one of the patterns from each group on the 7th day was picked and examined by the film adhesion test based on ASTM-3359B (Figure 3F-I) and Nie's method, in which nice and smooth edges were observed, which meant a 5B adhesion rating in regard to the ASTM-3359B standard.[5]

The stability of the printed Au electrodes was further examined by cyclic voltammetry (CV). The CV setup is shown in Figure 3N. Figure 3O shows the CV curves at different scan rates, and Figure 3P shows an isolated curve from Figure 3O, thereby showing how the 50-cycle stability test proceeded at each scan rate. From the CV test, it was shown that the Au electrodes had sufficient stability since the curves in any of the 50 cycles ran smoothly on the same tracks and the deviation of each data point in a cycle was less than 0.1 mA from the data points of other cycles.

### 2.3. Printed Solid-State Ion-Selective Nitrate Sensor

The demand for inexpensive sensors for nitrate sensing has been high since coated wire electrodes were introduced nearly 50 years ago. These electrodes paved the way for current developments in solid-state ion-selective electrodes (ISEs). ISEs convert the activities of adsorbed ions (analytes) into changes in electrical potential under high-impedance (near-zero current) conditions and can be used for rapid and quantitative ion sensing. By eliminating the internal filling solution, the fabrication of solid-state ISEs can be simplified and miniaturized. Most of the ISEs reported to date are based on glassy carbon disk electrodes, and their preparation is costly and laborious. There are only a few examples of ISEs fabricated by inkjet printing onto flexible substrates, and most of these are based on paper. Here, we report a method for the inkjet printing of Ag electrodes onto Kapton, followed by the deposition of polymeric membranes that enable them to function as solid-state ISEs with selectivity for the nitrate ion, $NO_3^-$.

In this study, inkjet-printed Ag electrodes and electrodes coated with ion-selective membranes (ISMs) are shown in Figure 4A(d) and (e), respectively. The ISMs were composed of poly(vinyl chloride) (PVC) and dibutyl phthalate (DBP) and doped with tetraoctylammonium nitrate (TOAN), a lipophilic salt that supports anion exchange.

Basically, the printed ISEs respond to changes in ionic concentrations by adjusting their electrical potential, which can be described with the Nernst equation:

$$\mathrm{EMF} = \mathrm{E^o} + \frac{2.303\mathrm{RT}}{\mathrm{z_iF}}\log \mathrm{a_i}$$

where EMF is the electromotive force (measured potential), $E^o$ is the potential constant, R is the ideal gas constant, T is the temperature, $z_i$ is the ion valence, F is the Faraday constant, and $a_i$ is the ionic activity.[23–29]

**Figure 4B** shows that a nearly Nernstian response of 56.6 mV/dec was observed with a linear sensitivity range between $10^{-4.5}$ M and $10^{-1}$ M, corresponding to a limit of detection of 2 ppm for $NO_3^-$. These results established that despite the low cost of materials and the simplicity of the fabrication process, inkjet-printed Ag electrodes could provide high-quality potentiometric readings. Thus, these electrodes have potential for use in agricultural and environmental applications.

## 2.4. Biomolecule Immobilization and Testing Performance for $Hg^{2+}$ Biosensor Fabrication

Here, we examined the application of specific inkjet-printed gold electrodes coupled with ssDNA for mercury sensing. **Figure 5A** shows the working electrode characterization. First, CV experiments were recorded at increasing scan rates to evaluate the interfacial interaction between the labeled electrolyte and the bare electrode surface. The results presented in Figure 5A(a) confirmed the reversible nature of the governing phenomena at the surface of the electrode, which was in line with the linear correlation obtained between increasing scan rates and current response ($r^2 = 0.9914$) and served as the confirmation of the purely diffusive mechanism governing the interface.

The layer-by-layer (LbL) assembly approach that was used for fabrication was characterized by electrochemical techniques. The results are summarized in Figure 5A(b) and (c), offering evidence of successful surface modifications with this fabrication process. The CV experiments presented in Figure 5A(b) show confirmation of the reduced current response during fabrication as an effect of the immobilization of nonconductive layers (short ssDNA and proteins) on the electrode surface. The above results were in full agreement with the PEIS plots presented in Figure 5A(c) by tracking down the changes upon transfer resistance ($R_{CT}$) fitted from the obtained Nyquist plot. These results agreed with the proposed immobilization mechanism, in which the printed gold layer corresponded to a face-centered-cubic (FCC) packing structure (Figure 5A(g)) that interacted with aptameric sequences via a strong thiolate-Au bond; these bonds created a self-assembled monolayer (SAM), where the chemisorbed radicals moved to a stable bridge-FCC position as proposed by Pensa et al.[30] and illustrated in Figure 5A(h). Finally, a layer of BSA was immobilized to block the surface (Figure 5A(i)), thereby avoiding any undesired nonselective interaction during testing.

The as-fabricated printed electrodes (Figure 5A(d)) were used as working electrodes (WE) in a conventional 3-electrode setup, having a commercial reference electrode (Ag/AgCl, RE) and a counter (Pt) electrode (CE) as illustrated in Figure 5A(e). For all experiments, $[Fe(CN)6]^{3-/4-}$ was used as a redox label to analyze the oxidation-reduction potential (ORP) and its correlation with the fabrication process and detection ability (Figure 5B(f)).

The proposed printed biosensing platform followed the well-described thymine-mercury-thymine detection mechanism,[31] by inducing a “hairpin-like” conformational change in the presence of mercury ions. During the detection event, a bridge was generated between the thymidine residues, forming a “base pair”, as illustrated in Figure 5B(a). As a result, the ions captured by the aptameric strands led to a change in the electrical properties of the working electrode and modified the interfacial resistance (Figure 5B(b)). The fabricated printed electrode was incubated with increasing concentrations of mercury solution ranging from 0 to 100 ppm, and the impedimetric response was acquired. The Nyquist plot presented in Figure 5B(d) served as evidence of the directly proportional relationship between the resistance response ($R_{CT}$) and target concentration, which was in full agreement with the proposed detection mechanism. The analytical performance of the proposed platform is presented in Figure 5B(e), showing a linear correlation ($r^2 = 0.8806$) between the $R_{CT}$ values (z-fitting the Nyquist plot) and increasing concentrations of mercury, as illustrated in Figure 5B(f).
Figure 5C shows the XPS survey spectra (Figure 5C(b)) for the DNA/BSA sample. The P 2p signal confirms the presence of the DNA aptamer, as shown in the inset. The signals also confirm the presence of C, N and O from the nucleotides and amino acids attached to the gold electrode. Core-level XPS data are shown in Figure 5C(c-f). These spectra corresponded to the high-resolution XPS scans, and the core-level energies of the components resulting from the fitting procedure are shown in **Table 1**. Despite the complexity of the chemical environment, the experimentally determined binding energies are in reasonable agreement with a previously reported paper.[32] The label assignment of each component is described as follows: (c) Au $4_{f7/2}$ single peak located at 83.5 eV was assigned to pure metallic gold, and no extra species were observed; (d) N 1s peak was composed of four signals and they were related to amine species (N1 to N3) and the amide bond (O=C-N) group (N4); (e) the O 1s core level was also decomposed into four components in good agreement with C-O, O-C=O, C=O and OH- species (O1 to O4); and finally, (f) the C 1s peak was fitted with the minimum number of components and they corresponded to C-C/C-H, C-N-C/C-O-C, N-C=O and O-C=O groups (C1 to C4). These results provided a successful confirmation of peptide bond formation. Au 4f was selected as a reference for energy calibration. Since surface electron compensation was applied during the measurement, most of the peak widths (FHWM values) were due to instrumental broadening in most cases.

### 2.5. Portable Protein Resistor Sensor

The portable protein resistor sensor is composed of a PCB sensor (**Figure 6D**) and a printed Ag resistor (Figure 6E). In this study, the Advanced Design System (ADS) was first used to simulate the performance of the PCB sensor. ADS allowed for the components of the bridge to be implemented in a way that modeled the system output when a sensor had a variable resistance. The system was then realized according to the schematics detailed in Figure 6(A). The construction of sensor 1 to sensor 2 was designed for different sensing ranges. To modify the sensing range of the bridge from sensor 1 to sensor 2, the value of R5 must be updated to match the expected value of the sensor. Figure 6(C) contains a CAD image of the PCB sensor with labels that correspond to the components that must be added to implement the sensing system. A voltage regulator was used to manage the battery voltage provided to the board (U1). An analog to digital converter (U3) was used to record the output of the Wheatstone bridge. The PCB also contained two different capacitance sensors (U4 and U5) to expand the sensing capabilities. Additionally, a Nordic system on a chip (SoC) was included on the board for data processing and communication. While these sensors have not been implemented in this work, there is potential for them to be utilized in future tests.

Figure 6(B) Sensor 1 displays the output of the system over a given resistance range when the system is matched to a resistance of approximately 20 ohms. As the system only worked within a limited range of resistance values, the system should be adjusted to closely match the expected output to accurately return a result, which is the reason why Figure 6(B) Sensor 1 shows a nonlinear response in some ranges.

After the PCB sensor was calibrated, a series of well-defined printed Ag resistors with a resistance of up to 16 ohms were used to indicate the accuracy of the PCB sensor. These Ag resistors were measured by the PCB sensor with replicates, and the results were compared with that of a multimeter. Figure 6F shows that the PCB sensor measurements were exactly the same as the multimeter measurements, and the deviations from the curves were simply from the differences of each resistor. Then, 0.1, 1, and 10 % BSA were used to create a blocking layer on top of the resistors, and the results are shown in Figure 6(G). The blocking affected the resistors with a resistance above 4 ohms and a resistor segment of less than 1 mm. Thus, the results showed a decrease in resistance after BSA blocking, which fit the report presented by Saab et al.[33] for simple biosensing with resistors.

**2.6. Organic Thin Film Transistor (OTFT) Fabrication**

In this study, a printable multilayer OTFT was fabricated on a Kapton NH film. The layers included a gate layer composed of Ag, a gate dielectric layer composed of crosslinked poly-4-vinylphenol (cPVP), a source (S)/drain (D) layer composed of Ag, and an organic semiconductor (OSC) layer composed of poly(triaryl amine) (PTAA). **Figure 7A(a-g)** schematically shows the construction of the printable organic TFT, and Figure 7A(h-k) shows the composition of each layer. Figure 7A(l-n) shows the images during the construction of the printable TFT samples. Figure 7A(l) shows the dimension of the first silver layer, Figure 7A(m) shows the dimension of the gate dielectric layer, which is constructed by masking and spin-coating procedures, and Figure 7A(n) shows the dimensions of the source and drain electrodes along with the area of the OSC layer. Figure 7A(o) further magnifies a part of the OSC area from Figure 7A(n). The magnified image shows that the OSC had a rainbow-like color. Figure 7A(p) shows the OTFT appearance. Figure 7A(q) shows the SEM image taken from the focused ion beam (FIB) sample preparation for a TFT cross-section. In Figure 7A(q), the angle was 52 degrees vertically for the image. Thus, the vertical length equaled the measured length divided by 0.788. The cross-section sample, which consisted of two silver layers, a dielectric and an OSC layer, was further investigated in Figure 7B by TEM.

The FIB lift-out TFT sample was first studied with low-magnification TEM in Figure 7B(a-c) with scale bars of 1 μm, 500 nm, and 200 nm. In these low-magnified images, the stacking layers, including the Kapton substrate, the Ag gate layer, the cPVP dielectric, the Ag source layer, and a thin OSC layer, were clearly observed. The high-angle annular dark-field (HAADF) STEM image was further used to investigate the FIB lift-out TFT sample, as shown in Figure 7B(d-f). The STEM images offered another way to observe the FIB lift-out TFT sample with different contrasts. Furthermore, the electron-dispersive X-ray spectroscopy (EDS) mapping for C, Ag, and Pt with the original HAADF-STEM images in Figure 7B(g-i) provided a second proof for the existence of each layer in the OTFT. The elemental profiles from the top to bottom are displayed in Figure 7B(j) for the EDS line scan, which also clearly examines the layers in a sequence of Ag (gate), cPVP, Ag (source), and OSC.

Figure 7C shows the XPS spectra of each step-by-step deposited layer to obtain the OSC/Ag/cPVP/Ag architecture. Since the first layer is simply Ag, as shown in Figure 7C(a), the survey scan spectra showed the presence of all elements with a strong Ag 3d signal. From the high-resolution XPS spectra, Ag $3d_{5/2}$ (Figure 7C(c)) was composed of two signals associated with Ag(0) and Ag(I) components. The silver was sensitive to being easily oxidized when in contact with an air atmosphere, and this result was quite logical for this kind of procedure. However, metallic silver continued to be the main species found on the top surface. The C 1s (Figure 7C(d)) peak was assigned to two components associated with the main aliphatic signal and C-O species. The O 1s (Figure 7C(e)) peak was composed of several peaks related to carbonyl, carboxyl and related species in the same way that was described in Figure 5C. Each of these components was assigned to the chemical groups present in the Kapton substrate. The former was corroborated in the N 1s peak (Figure 7C(f)), where at least two components were selected after fitting. For the second layer (cPVP, Figure 7C(g)), the survey scan spectra showed the C, O and N core levels, but no silver signal was detected. This result could be associated with the fact that the layer formed by poly(melamine-co-formaldehyde) methylated (PMFM) and poly(4-vinylphenol) (PVP), which are both labeled cPVP, was wide enough to inhibit the XPS detection of the silver underneath. This result was indicative that the cPVP crosslinking layer was successfully formed by attaching the melamine ring-terminated hydroxyl groups (O-R) with phenol groups, as was confirmed by the results of the high-resolution XPS data collected for the C 1s (Figure 7C(i)), O 1s (Figure 7C(j)) and N 1s (Figure 7C(k)) core levels. A surface charge effect could be present because all peak envelopes were anomalously wide and the binding energies were clearly shifted from the expected values (± 1 eV). This result allowed us to conclude that most of the peak width due to instrumental broadening was increased for positive charges along the surface as consequence of the low electrical performance of the cPVP layer compared with that of the metallic silver. Note that the work function depended on both the spectrometer and the material, and no surface electron compensation was applied because the XPS data should be comparative for all analyzed layers. For this reason, a specific discussion about the nature of each PE-peak signal was omitted since the fitting procedure was clearly affected by the instrumental resolution. For the third layer, the survey scan spectra (Figure 7C(l)) confirmed the presence of the Ag, C, N, and O core levels. Regarding the high-resolution spectra (Figure 7C(m)), the silver signal behaved in the same way as the first layer, but the peak shoulder associated with the oxidized species was less evident; therefore, the Ag(0)/Ag(I) ratio was higher. The C 1s (Figure 7C(n)) and O 1s (Figure 7C(o)) core-level spectra confirmed the previous discussion about surface charge compensation. For both cases, the peaks were well resolved into their components, the peak envelopes were symmetrical, sharp and narrow, and the binding energy positions were coherent with those of the cPVP layer, indicating that the silver spots of the third layer were thin enough to allow elemental detection of the layer underneath. Nevertheless, the high-resolution XPS data from these core levels were less intense, as expected, and the N 1s signal was too weak to be decomposed. Finally, in the fourth layer (Figure 7C(p)), the OSC layer corresponded to the p-type semiconductor poly(triaryl amine) (PTAA), which substantially improved the open-circuit voltage and fill factor of the OSC. The survey scan spectra (Figure 7C(q)) showed only the C, N and O signals, and no silver was detected. This result means that the 20-30 nm PTTA layer was wide enough to block most of the ejected photoelectrons from the surface of the third silver layer. For the C 1s (Figure 7C(r)) core level, the signal was decomposed into two main components, which were associated with aliphatic C-C/C-H and N-C=O bonds. For the O 1s (Figure 7C(s)), the peak was composed of several peaks associated with C-O from the cPVP layer underneath and the adsorbed C-O/O-H species from the environment. The weak N 1s signal was not able to be properly decomposed. Hence, the overall XPS results confirmed the successful formation of the OSC/Ag/cPVP/Ag circuit. Figure 7D further shows the electrical performance of the printed TFTs. Figure 7Da shows the typical I-V output characteristics for a TFT with an optimal printing parameter, a W/L ratio that equaled 90. The printed TFT was functional and demonstrated operational stability for weeks. Figure 7D(b) shows the variation in the conductance channel for the linear, pinch-off, and saturation regimes corresponding to the I-V output characteristics in Figure 7D(a). The band diagrams in Figure 7D(c) explained the flat band status and the band bending when the negative gate voltage was applied in Figure 7D(a), thereby leading to the onset of the accumulation mode and the increase in drain current, $I_D$. Furthermore, Figure 7D(d) shows the extraction of contact resistance $R_c$, carrier mobility for the linear regime, $\mu_{lin}$, carrier mobility for the saturation regime, $\mu_{sat}$, and threshold voltage, $V_{th}$, from the transmission line method (TLM) and the transfer curves inserted in the subgraphs (I-IV). Figure 7D(d) (I) first shows the extraction process of $R_c$ from 20-TLM sets of TFTs and obtained the average channel width-normalized contact resistance as $0.48 \pm 0.11$ MΩ·cm. Although the TFT was printed with an optimal W/L printing parameter, differences were still observed within different printed batches due to the deviation of the printer mechanism and the variation in the actual channel lengths. Thus, histograms for the extracted parameters must be shown in the following discussion due to their variation. Figure 7D(d) (II-III) shows the extraction and variations of $\mu_{lin}$ and $\mu_{sat}$ from transfer curves $I_D$-$V_G$ and $I_D^{0.5}$-$V_G$, respectively. Their histograms showed the average values of $\mu_{lin}$ and $\mu_{sat}$, which were equal to 0.0083 $cm^2V^{-1}S^{-1}$ and 0.0237 $cm^2V^{-1}S^{-1}$, respectively. Figure 7D(d) (IV) further presents the threshold voltage, $V_{TH}$, as an average value of -6.75 V, which was extracted from the transfer curves $I_D^{0.5}$-$V_G$.

## 3. Conclusion

In conclusion, to meet the current technological needs of low cost, simple fabrication, and high stability, 40% Ag and 25% Au inks were matched with Kapton NH substrates to fabricate sensors that met the requirements of biochemical reactions in aqueous and organic solutions and to construct OTFTs that had stable output, sufficient integrity, and long-lasting performance. The stability of the resulting platforms was examined by immersion and sonication in solvents and an adhesion test based on ASTM-3359B. After confirming the stability of both of the resulting metal-Kapton substrates, four different novel applications from our research groups were applied. First, an on-demand detection of $NO_3^-$ for use in the agriculture industry. We reported an inkjet-printed nitrate sensor that adopted a type of solid-state ion-selective electrode, which was composed of inkjet-printed Ag electrodes and a layer of ion-selective membranes (ISMs). The ion-selective nitrate sensor displayed a Nernstian response of 56.6 mV/dec and a linear sensitivity range between $10^{-4.5}$ M and $10^{-1}$ M with a limit of detection of 2 ppm $NO_3^-$. Thus, this sensor shows promising selectivity and sensitivity for the detection of nitrate ions and reveals its potential application with other analytes. In addition, we also reported a cost-efficient solution for an industry pollutant, $Hg^{2+}$, which continues to create crises in our environment. This fully scalable inkjet-printed mercury sensor adopted a flexible Au substrate to provide a perfect location for a highly selective ssDNA sequence. The hairpin-like conformational changes of the ssDNA in the presence of mercury ions provided efficient real-time monitoring for the pollutant that needs to be urgently taken care of. We demonstrated a linear correlation ($R^2$ = 0.8806) between the $R_{CT}$ values (z-fitting the Nyquist plot) and increasing concentrations of $Hg^{2+}$. A further optimization of the printed Au patterns improved the LOD to 0.005 ppm. Furthermore, a portable PCB sensor paired with inkjet-printed Ag resistors was reported as a resistance sensor and provided a simple method for protein sensing. The PCB sensor provided the same accuracy as an unportable digital multimeter and read the slight changes of resistance from the printed Ag resistors. Thus, this sensor exhibited a potential application of functionalized circuits. Moreover, a low-cost and long-lasting inkjet-printed p-type organic thin film transistor, as a major component in integrated circuit fabrication, was also reported. The output characteristics were presented with gate voltages from -5 to -20 V. The average normalized contact resistance, which was extracted by the transmission line method, was 0.48 ± 0.11 MΩ·cm. The mobility in the linear and saturation regimes was extracted from the transfer characteristics and had values of 0.0083 $cm^2V^{-1}S^{-1}$ and 0.0237 $cm^2V^{-1}S^{-1}$, respectively. The average threshold voltage was also extracted from the transfer characteristics and had a value of -6.75 V. Overall, this sensor provided accurate positive/negative responses, and the printed OTFT exhibited durable and stable performance for weeks. These reported applications fully demonstrated the merit of the special 25% Au and 40% Ag inkjet-printed platforms on Kapton NH films and reveal their potential for being applied in the biosensor and integrated circuit industries.

**4. Experimental Procedure**

*Materials and Electrode/Circuit Fabrication:* Various raw materials, including inkjet printing inks and flexible polymer substrates, were purchased from Novacentrix (Austin, TX, USA), UT Dots, Ink (Champaign, IL, USA), and DuPont (Wilmington, DE, USA). Materials: 40 wt% silver ink JS-B40G was purchased from Novacentrix (Austin, TX, USA), which required a curing temperature over 180 °C, and 25 wt% gold ink UTDAu25IJ was purchased from UT Dots, Ink (Champaign, IL, USA), which required a curing temperature over 200 °C. The Kapton HN film with a thickness of 125 μm was selected as the substrate due to its wide application range in regard to temperature and more stable performance at higher temperatures (-269 to 400 °C) with different types of Kapton. Dimatix Materials Printer DMP 2850 (Santa Clara, CA, USA) and 10 μl cartridges were used for printing electronic traces. A professional hot plate from VWR (Batavia, IL, USA) was adopted for conductive ink sintering. In the process of inkjet-printed electrode fabrication, first, a letter-sized Kapton HN film was placed on the Dimatix platform with the vacuum system turned on. Second, the plate temperature was set to 50 °C to control the contact angle of the drops lying on the substrate, and the cartridge temperature was set to 38 °C to maintain the flowability of the inks. Third, various drop spacings and printing resolutions were applied to optimize the printed patterns. After the printing was complete, the finished patterns were dried at room temperature for 10 min and then moved to a preheated hot plate for a temperature survey from 190 °C to 400 °C. Next, optimal temperatures of approximately 330 °C to 400 °C for Ag and 250 °C to 330 °C for Au were chosen for the following applications, and in each application, the printed patterns were maintained on a hot plate for 10 min to finish sintering.

*General Analyses for the Printed Thin Films:* TEM images of the two kinds of inkjet printing ink were acquired with an FEI Tecnai G2 20 transmission electron microscope (ACC V:200 kV, FEI Corp., Hillsboro, OR, USA). SEM images of the adhesion tests were acquired with an FEI Quanta 3D FEG scanning electron microscopy (magnification 200x, 5 kV, FEI Corp., Hillsboro, OR, USA). The thickness and surface morphology analysis was obtained by using a KLA-Tencor Stylus Profilometer (Milpitas, CA, USA).

*Thin Film Mechanical Property Simulations and Measurements:* The finite element model was developed in the commercial package ABAQUS/Standard to investigate the gradients in the linear-elastic axial $\epsilon_{xx}(x, y)$ strain field from the tensile and bending deformation of the coated films, and the results were compared with the results from the actual experiments. To further characterize the mechanical properties of the Au and Ag thin films, a nanoindentation test was performed on the samples (in the supplementary file).

*Solid-State Ion-Selective Electrodes (ISEs) for Nitrate Sensors and the Detection Process:* High-molecular weight PVC (Selectophore grade, Sigma-Aldrich), DBP (99%, Alfa Aesar), and TOAN were used as the polymer matrix, plasticizer, and anion exchanger, respectively, for the ISM. Tetraoctylammonium nitrate (TOAN) was prepared from TOAB (98%, Sigma-Aldrich) by ion-exchange chromatography using Amberlite IRA-400 (OH) resin beads; the endpoint for anion exchange was determined by AgBr precipitation. Tetrahydrofuran (THF) was obtained from Fisher Scientific, and all aqueous solutions were prepared using deionized (DI) water.

PVC (2 g), DBP (4 g), and TOAN (0.134 g) were dissolved in THF (19 mL), drop cast in 0.1-mL aliquots onto one end of the inkjet-printed Ag electrodes and then dried in air. The final nitrate-selective membranes were comprised of 2.1% TOAN, 32.6% PVC and 65.3% DBP. The remainder of each Ag electrode was passivated by drop casting a silicone heat-cured RTV adhesive coating (Silicone Solutions Inc.) to minimize the interference of the aqueous electrolyte on the potentiometric measurements, which were collected using a multichannel data acquisition system (National Instruments, PXI-6225). A double-junction Ag/AgCl electrode (Orion (900200 Sure-Flow) with an outer filling solution of 1 M lithium acetate was used as a reference.

A total of 13 inkjet-printed Ag electrodes were coated with the ISM and silicone solutions and left to dry under ambient conditions for 24 hours prior to use. The ISEs were placed inside a 3D-printed holder (Figure 4A(b)) and conditioned in 1 mM $KNO_3$ for a minimum of 10 hours. Potentiometric responses were recorded using aqueous $KNO_3$ solutions ranging from $10^{-5}$ to $10^{-1}$ M in order of increasing concentration. Electrodes were rinsed in DI water between each measurement.

*Printed Electrode Cleaning and Aptamer Immobilization for the Electrochemical $Hg^{2+}$ Sensors:* A cleaning step was performed on all printed Au electrodes by sonication with ethanol and DI water for 2 min; voltammetry cycles (scan $E_{WE:}$ +0.6 to −0.7; scan rate: 80 mV/s) were applied until a stable plot was obtained.

The customized thiol-modified DNA aptamer was synthesized by IDT Integrated DNA technologies (Coralville, IA, USA) with a sequence of 5′thiol/MC6-D/TT TCT TCT TTC TTC CCC CCT TGT TTG TTT 3' immobilized on the gold electrode surface via a stable thiol-Au bond using the method described in our previous publication with a few modifications.[32] Briefly, the aptamer sequence was activated by excess TCEP (100: 1) before use. Then, 10 μl (10 μM) of the aptamer was drop casted on the printed gold electrode, and the activated aptamer was incubated on the electrodes for 2 h at 25 °C for aptamer immobilization. After incubation, a subsequent washing step was conducted with a phosphate-buffered saline (PBS) solution, which was suitable for cell culture and obtained from Sigma-Aldrich Co. (St. Louis, MO, USA) (10 mM, pH 7.4). Next, 10 μl of bovine serum albumin, BSA (10% PKL, Seracare, Milford, MA, USA) was used to block the reactive surface of gold. The sample was then finalized by being washed with DI water. The blocked and functionalized gold electrodes were stored in sterile DI water under refrigeration.

*Electrochemical Measurements for the Electrochemical Sensors:* Cyclic voltammetry (CV) and potential impedance spectroscopy (PEIS) were adopted for the electrochemical characterization of the electrode and $Hg^{2+}$ detection in the Faradaic mode at the open-circuit voltage (OCV). The supporting electrolyte, ferrocyanide (5 mM) and ferricyanide (5 mM) (1:1) $[Fe(CN)6]^{3-/4-}$, were mixed in 1 mM KCl solution and provided as reduction/oxidation reaction probes in the solution tank for the electrochemical experiments (details are in the supplementary file).

*Fabrication and Simulation of the Sensor PCB/Protein Measuring Process with the PCB Resistance Sensor:* A resistance sensor was realized using a Wheatstone bridge, which is a network of resistors that can be used to determine the value of an unknown resistor within a specific range, on a custom printed circuit board (PCB). The value of the unknown resistor was provided by the voltage differential at a specific point in the circuit. The PCB was a two-layer board with an FR-4 substrate and was manufactured by Advanced Circuits (Aurora, CO, USA). The PCB was manufactured according to the CAD image in the supplementary file. A simulation of a Wheatstone bridge was carried out using the Advanced Design System (ADS) electronic design software produced by Keysight Technologies (Santa Rosa, CA, USA). After the PCB sensor system was set up, a series of printed silver resistors with a resistance of up to 16 ohms were coupled with the PCB resistance sensor for protein sensing. In this study, BSA was used as the target protein. During the sensing steps, 5 µl of a BSA blocking solution was dropped on the circular central area of the printed resistors. The resistors were incubated with the BSA drop on top of the resistors for 5 min to create a blocking layer, slightly washed with deionized water, and then dried. The resistance measurements were taken by the PCB sensors before and after the blocking process.

*Materials and Fabrication Process for the OTFTs:* Commercial chemicals used in the organic thin film transistor fabrication, namely, poly(triaryl amine) (PTAA), crosslinked poly(4-vinylphenol) (cPVP), and others, were purchased from Sigma-Aldrich Co. (St. Louis, MO, US). A reported recipe[34] for TFT on PEN was referenced, and in this study, parts of it were applied to the Kapton substrate with some major modifications. Briefly, under magnetic stirring for 4 hours at room temperature, 0.9 g PVP with Mw 25000 was dissolved in propylene glycol monomethyl ether acetate (PGMEA), and under magnetic stirring for 6 hours, poly(melamine-coformaldehyde) methylated (PMFM) with 432 Mn was added to the PGMEA solution as a crosslinker. The solution was then filtered by a 0.2 mm syringe filter and stored at 25 °C before use. For the solution of the OSC layer, a p-type semiconductor PTAA was made with a mesitylene solution and stored at 25 °C before use.

Silver ink (40%) was applied to the Kapton substrate for the fabrication of a metal gate, and the printed traces went through a high-temperature sintering procedure of up to 400 °C to increase their conductivity. Then, a masking step was performed before the coating of the dielectric layer, followed by the cPVP coating using a spin coater at a speed of 500 to 1500 rpm on a 3750 µm × 3750 µm passivation region. The dielectric layer was dried at 25 °C for 1 hour and baked on a hot plate at 150 °C for 1 hour in a fume hood. After the dielectric layer was stabilized, the source (S) and drain (D) electrodes were deposited on top of the cPVP regions with a special low-temperature sintering procedure to avoid damage to the dielectric layer. After that, the S and D layers were stabilized, and the OSC layer was then deposited on top of the S and D layers using a spin coater at a speed of 200 to 500 rpm, followed by a baking process at 100 °C for 30 min.

*Thin Film Analyses and OTFT Characterization:* Briefly, the observation of thin film layers in TFT was acquired first by a focused ion beam (FIB) sample preparation on top of a source electrode region with an FEI Quanta 3D FEG scanning electron microscopy (magnification 200x, 5 kV, FEI Corp., Hillsboro, OR, USA), and the characterization of the TFT was tested by a Probe Station with a Keithley SCS4200 semiconductor parameter analyzer (Solon, OH, USA). In the cutting process, the TFT sample was treated by an FIB cut for scanning/transmission electron microscopy (S/TEM) analyses with the following procedure, in which the TEM sample was lifted out by an Omniprobe manipulator and thinned by a low-energy FIB on a Thermo Fisher Quanta 3D dual beam scanning electron microscope. The sample was first subjected to e-beam deposition at a thickness of 1 μm (3.4 nA current), which was performed on a gas injection system (GIS). Then, the samples were treated with ion-beam deposition, which formed a 6 μm thick (0.5 nA current) layer and was performed on a gas injection system (GIS). Then, a thinning process was conducted to reduce the FIB cut sample thickness from 2 μm to 400 nm (30 kV, 0.3 nA) and 400 nm to 120 nm (5 Kv, 47 pA). Next, a final cleaning was executed (2 kV, 27 pA), and the sample was subjected to S/TEM for TEM and high-angle annular dark-field (HAADF) imaging and elemental mapping. TEM analyses were conducted on a Thermo Fisher Talos 200X analytical microscope operated at 200 kV. Fischione high-angle annular dark-field (HAADF) detectors and super X electron-dispersive X-ray spectroscopy (EDS) detectors built in Talos 200X were used to perform STEM and EDS chemical mapping analyses.

*Surface Chemical Composition Analyses:* The chemical composition of the surface of the DNA/BSA and OSC/Ag/cPVP/Ag samples was determined by X-ray photoelectron spectroscopy in the NAP-XPS Laboratory facility at SIU-UdeA. This system provides a hemispherical PHOIBOS 150 1D-DLD electron energy analyzer and a μ-FOCUS 600 NAP X-ray monochromator (SPECS GmbH). The PE measurements were acquired at room temperature with photon energies of the Al Kα radiation source (1487 eV) operated at 13 kV and 100 W. The data were collected with a pass energy of 30 eV and a step size of 0.01 eV for high-resolution data recording and 90 eV and a step size of 1 eV for the survey spectra. The working pressure was kept at $5 \times 10^{-9}$ mbar, and charge compensation was achieved by a low-energy electron flood gun working at a 3 eV cathode voltage and a 2 μA emission current. The homogeneity of the samples was tested by observing several points on the surface, and no significant changes were observed after 20 scans. The HR spectra were fitted using a Gaussian-Lorentzian blend (GL 30%) and a Shirley-type background subtraction. The adventitious C 1s core-level line was selected as a reference to calibrate the energy scale.

**Conflict of Interest**

The authors have no conflicts of interest to declare.

**Figures**

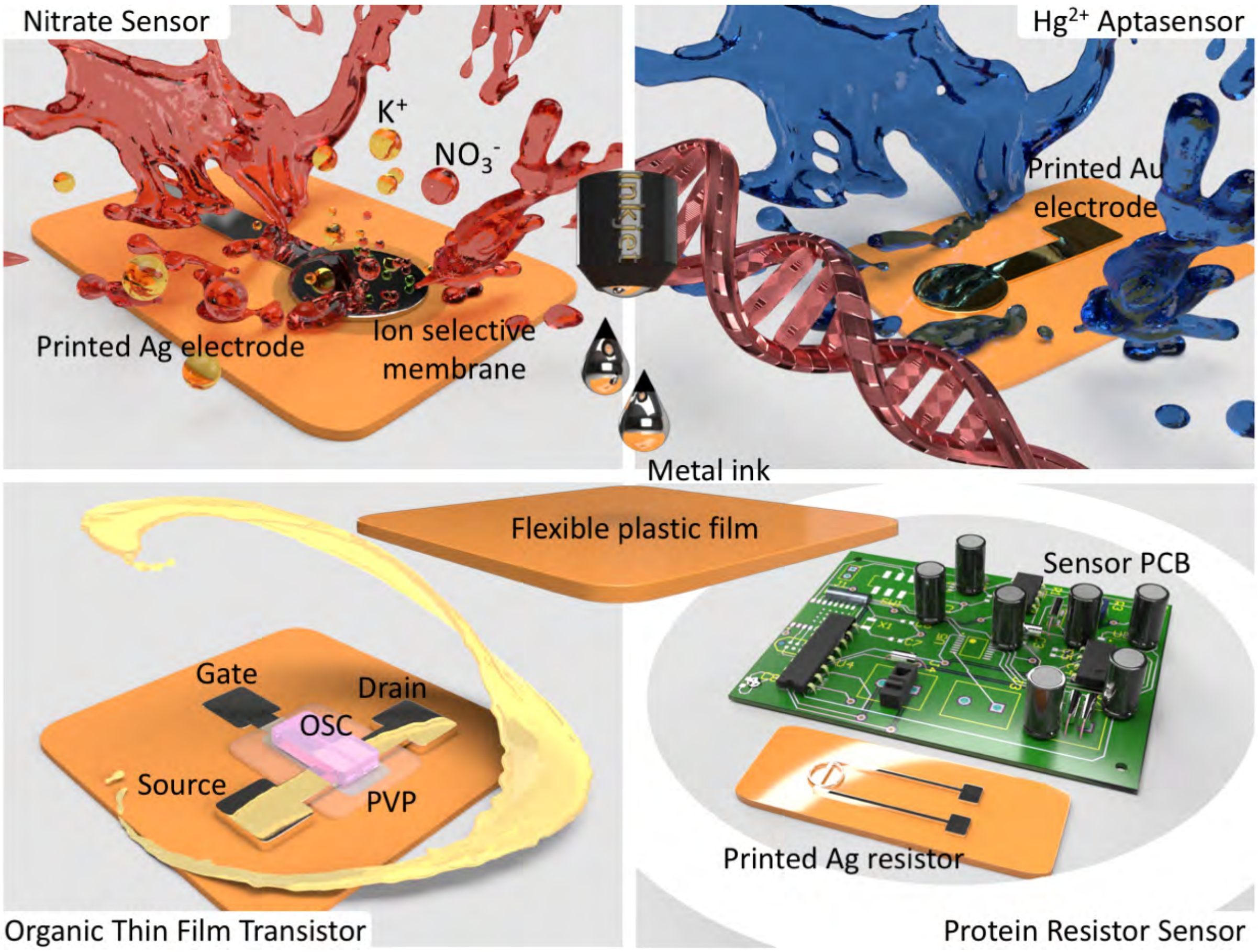

**Figure 1.** Schematic illustration showing the four different applications. (Top-left) Printed Ag nitrate sensor. (Top-right) Printed Au mercury aptasensor. (Bottom-left) Printed Ag organic thin film transistor. (Bottom-right) Printed Ag protein sensing system.

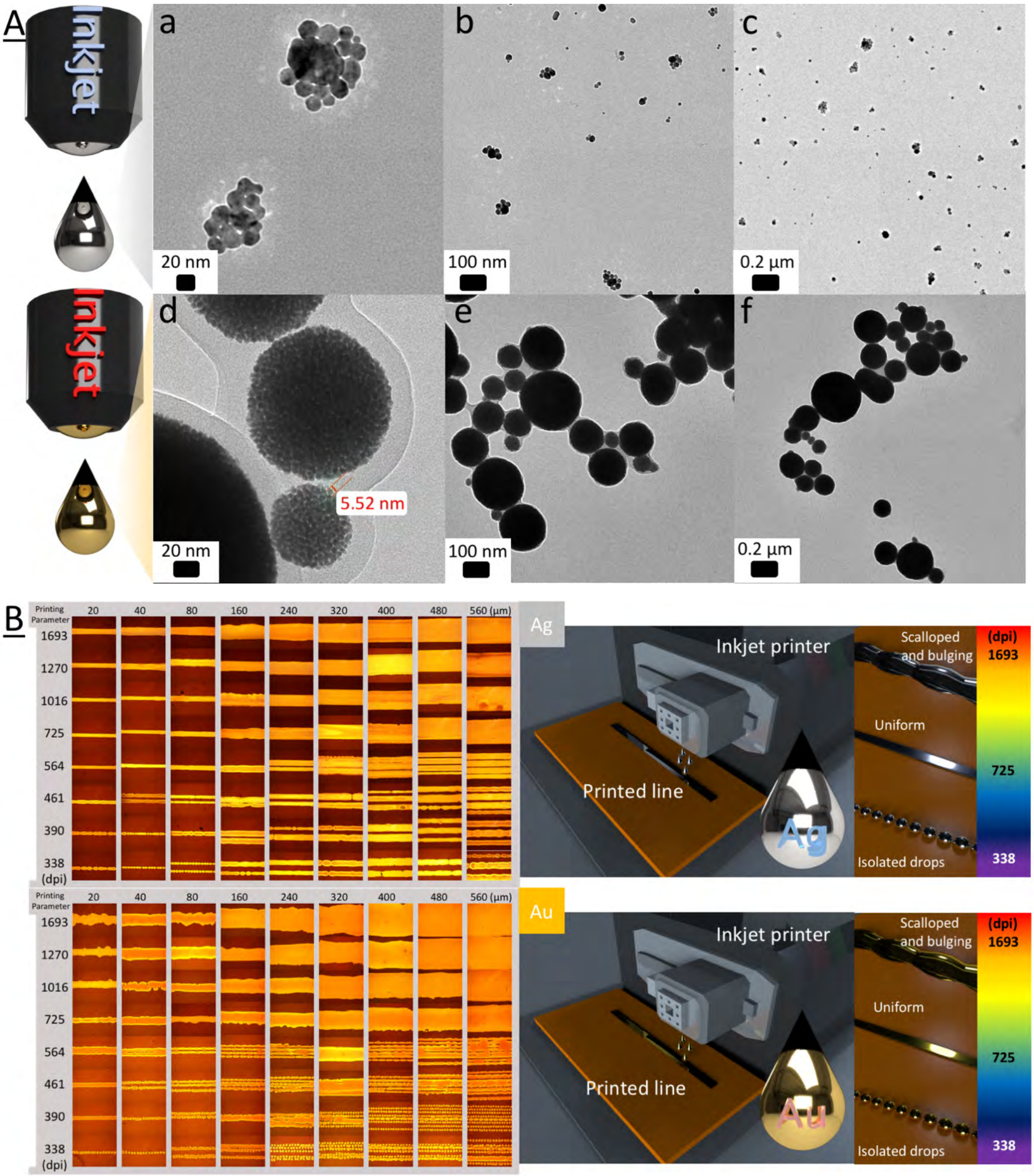

**Figure 2.** A) TEM images of the two kinds of inkjet printing ink: silver nanoflakes with a) 20 nm, b) 100 nm, and c) 0.2 μm scale bars and gold nanoparticles with d) 20 nm, e) 100 nm, and f) 0.2 μm scale bars. B) Resulting printing traces of both (upper) silver and (lower) gold ink with (y-axis legend) printing resolutions of 1693, 1270, 1016, 725, 564, 461, 390, and 338 dpi and with (x-axis legend) digital printing patterns with widths of 20, 40, 80, 160, 240, 320, 400, 480, and 560 μm on the Kapton substrate.

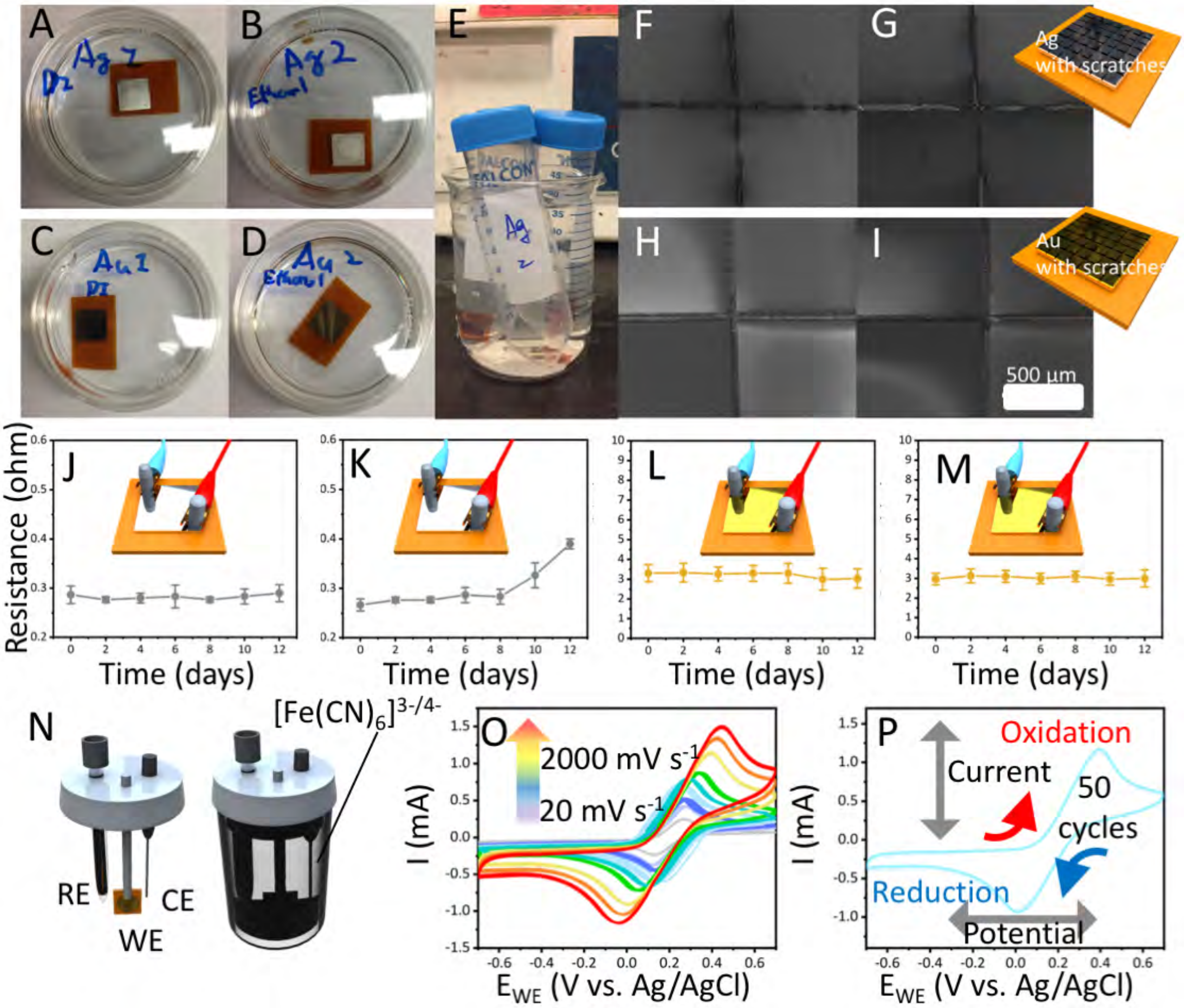

**Figure 3.** Photographs of the experimental setup: A) Ag pattern in deionized water, B) Ag pattern in ethanol, C) Au pattern in deionized water, and D) Au pattern in deionized ethanol; E) sonicated Au/Ag patterns in solutions; SEM images of the tape tests for the F) Ag pattern in deionized water, G) Ag pattern in ethanol, H) Au pattern in deionized water, and I) Au pattern in deionized ethanol, based on ASTM-3359B; and the resistance variation in the J) Ag pattern in deionized water, K) Ag pattern in ethanol, L) Au pattern in deionized water, and M) Au pattern in ethanol over time (days); N) experimental cyclic voltammetry (CV) setup for stability; and the CV curves at O) different scan rates and the P) electrode stability tests over 50 cycles.

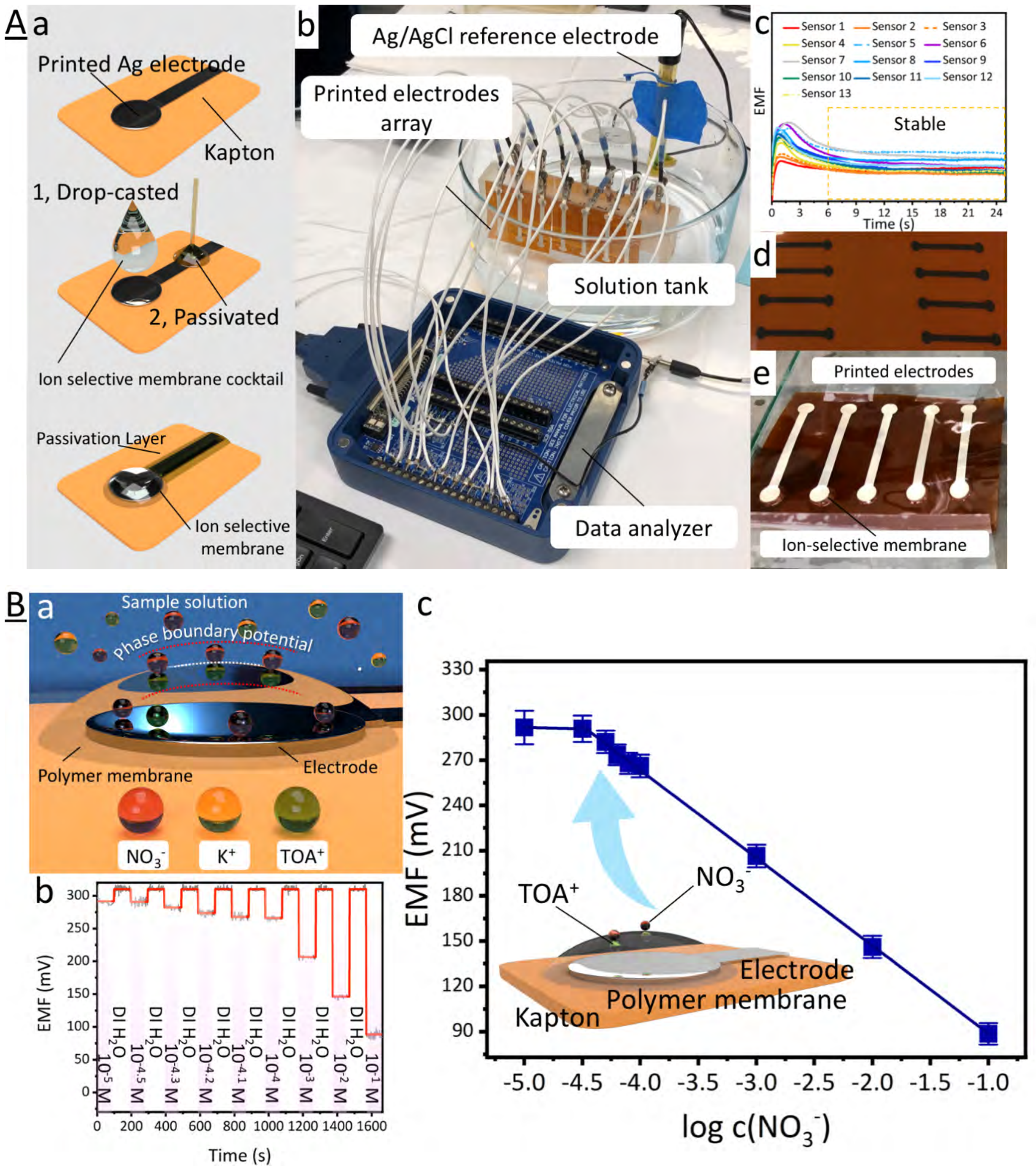

**Figure 4.** A: a) Stepwise fabrication of nitrate ISEs, b) potentiometric testing setup including multiple nitrate sensors paired with a commercial reference electrode in aqueous nitrate solution and the multichannel data acquisition system, c) 1-day conditioning data of 13 nitrate sensors in 1 mM nitrate solution, d) inkjet-printed Ag electrodes on Kapton, and e) complete nitrate sensors. B: a) Visual demonstration of how the phase boundary potential (EMF) is generated by oppositely charged ions facing each other in ISEs, b) real-time open-circuit measurements of a nitrate sensor at different nitrate concentrations (fitted curve is in red; real signal is in gray), and c) calibration curves and linear response of all nitrate sensors.

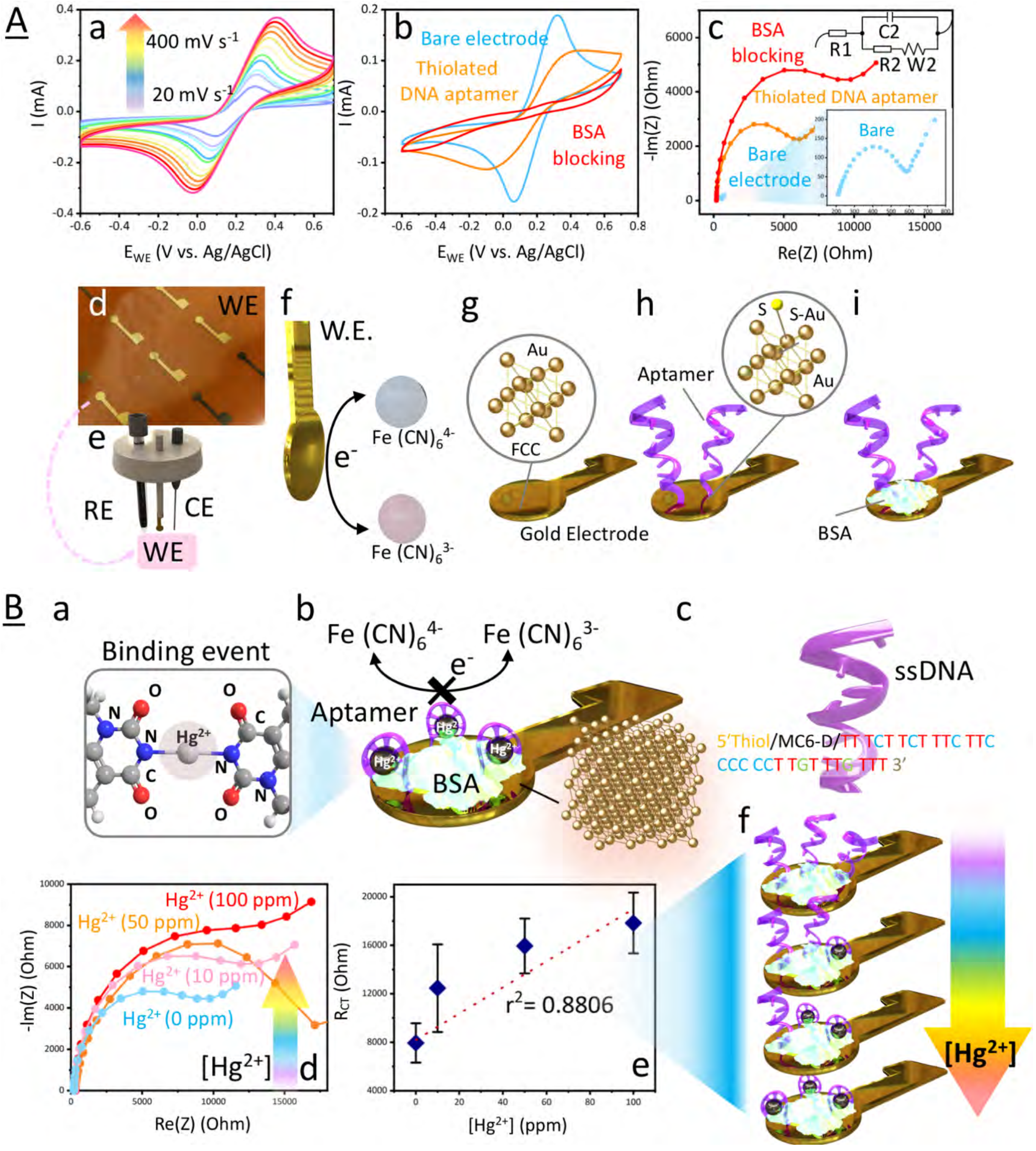

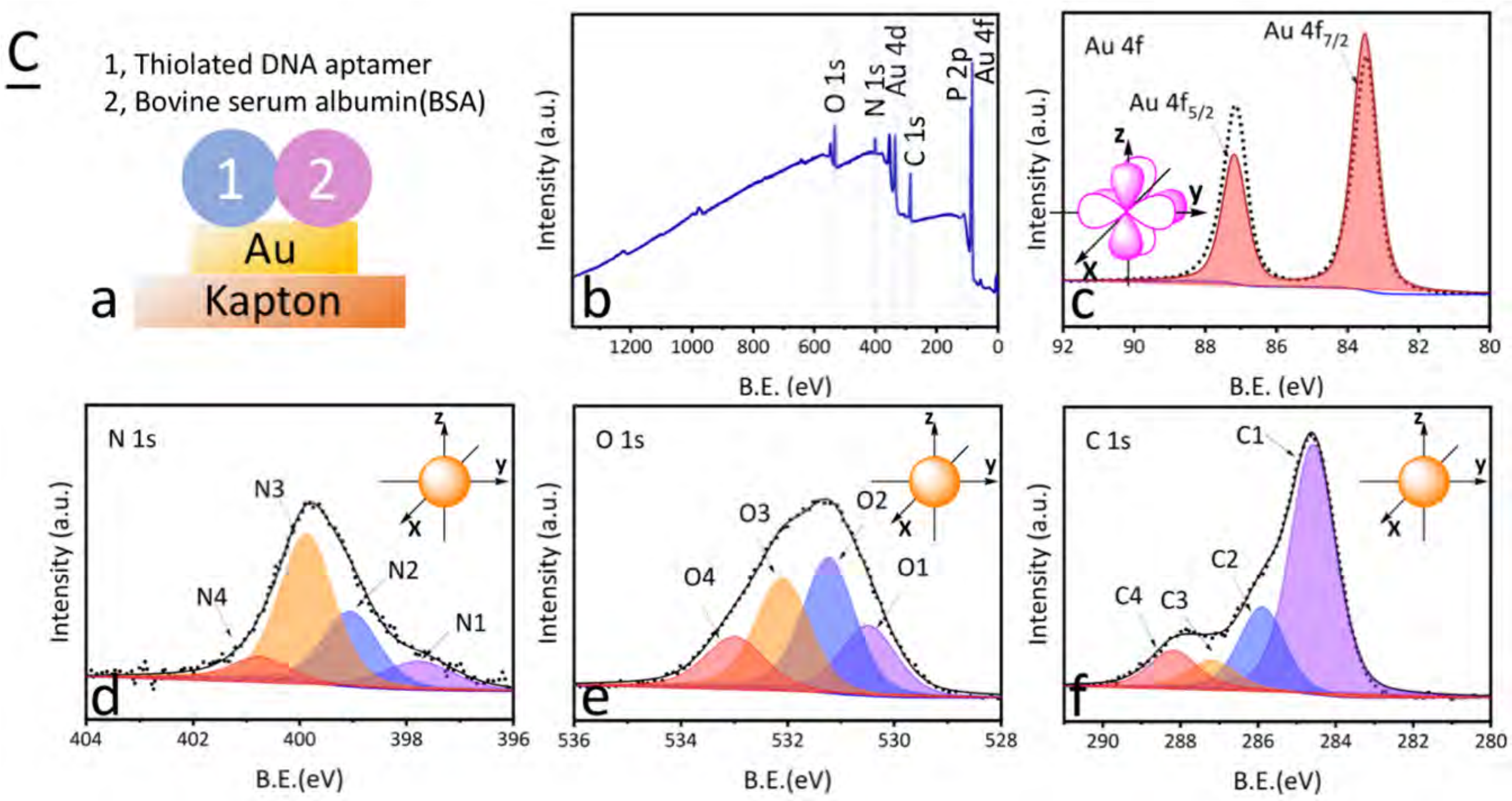

**Figure 5.** A) Working electrode fabrication and characterization: a) scan rate optimization, b) CV curve variation within the blocking and DNA immobilization steps, c) impedance variation within the blocking and DNA immobilization steps, d) appearance of Au electrodes and e) the setup as a working electrode in the CV tests, f) chemical mechanism in the CV tests, g) FCC structure of the Au electrodes, h) S-Au boding mechanism used for DNA immobilization, and i) BSA blocking. B) Platform performance, the detection of mercury ions in PBS: a, b) detection mechanism of mercury, c) sequence of the specific ssDNA for mercury detection, and d) CV curves for mercury detection at different concentrations. C) Schematic illustration of the DNA/BSA sample on Kapton (XPS scan spectra): a) survey spectrum b) and the decomposition of the high-resolution spectra for the Au 4f c), N 1s d), O 1s e) and C 1s f) core levels.

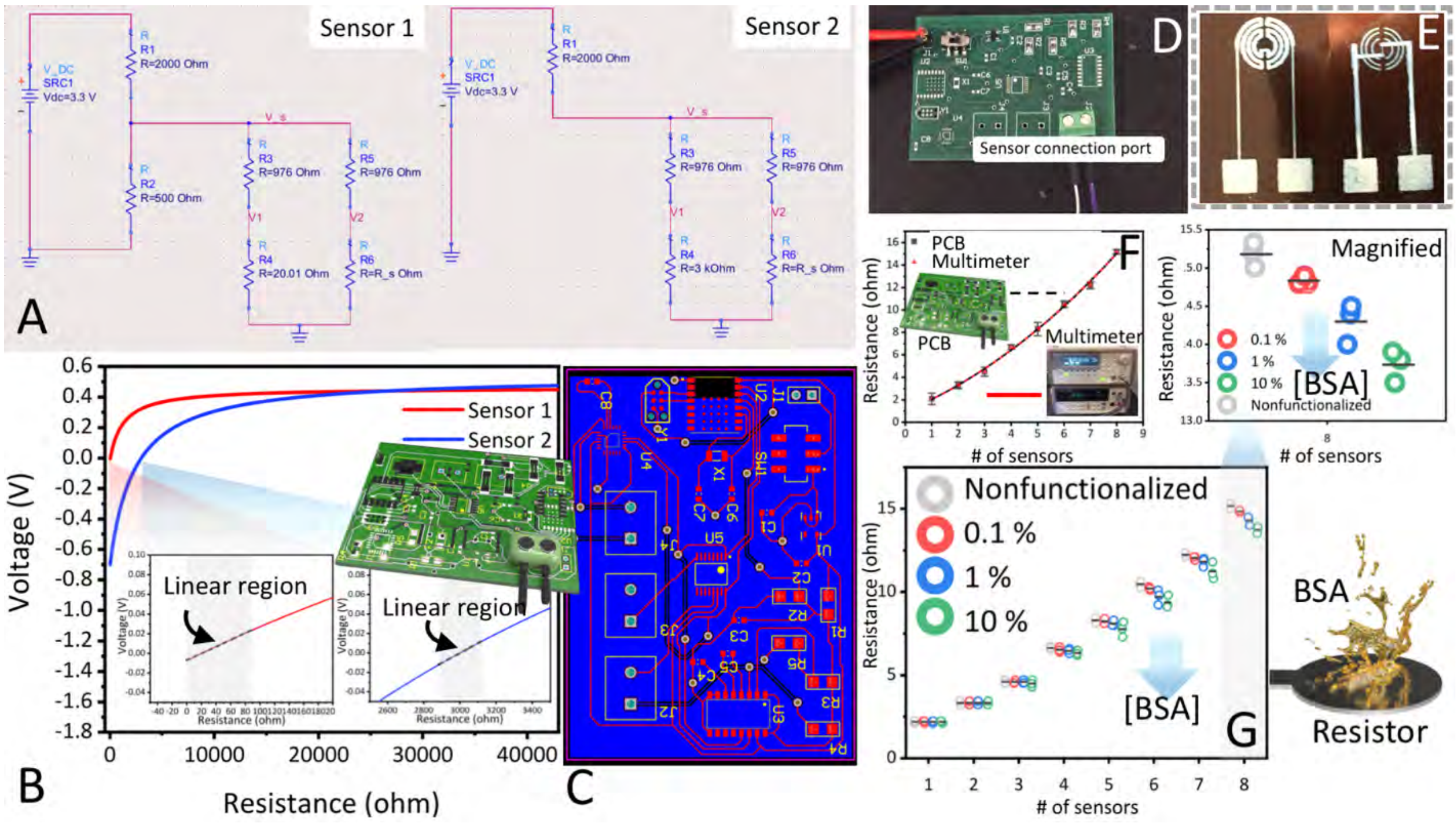

**Figure 6.** A) Advanced Design System (ADS) simulation schematics for two separate Wheatstone bridges matched to different sensor values. B) ADS simulation sensor voltage output for a varied sensor value. C) PCB sensor CAD image (J1 Header 1, J2 Header 2, J3 Header 3, J4 Header 4, SW1 power switch, R1 Resistor 1, R2 Resistor 2, R3 Resistor 3, R4 Resistor 4, R5 Resistor 5, C1 Capacitor 1, C2 Capacitor 2, C3 Capacitor 3, C4 Capacitor 4, C5 Capacitor 5, C6 Capacitor 6, C7 Capacitor 7, C8 Capacitor 8, U1 Voltage regulator, U2 NRF52 module, U3 NAU7803 module, U4 PCAP-04 module, U5 AD7747 module, and Y1 Debugger connection point.) D) Appearance of the applied PCB sensor. e) Appearance of the applied resistors. F) Comparison of the PCB sensor measurements and the multimeter measurements coupled with a series of resistors with the resistance from resistor #1 (avg. = 2.1 ohms) to #8 (avg. = 15.2 ohms). G) Comparison of the PCB sensor measurements for the resistor blocked with different concentrations of BSA (0.1 (red), 1 (blue), and 10 % (green)), in which the circles indicate the replicates.

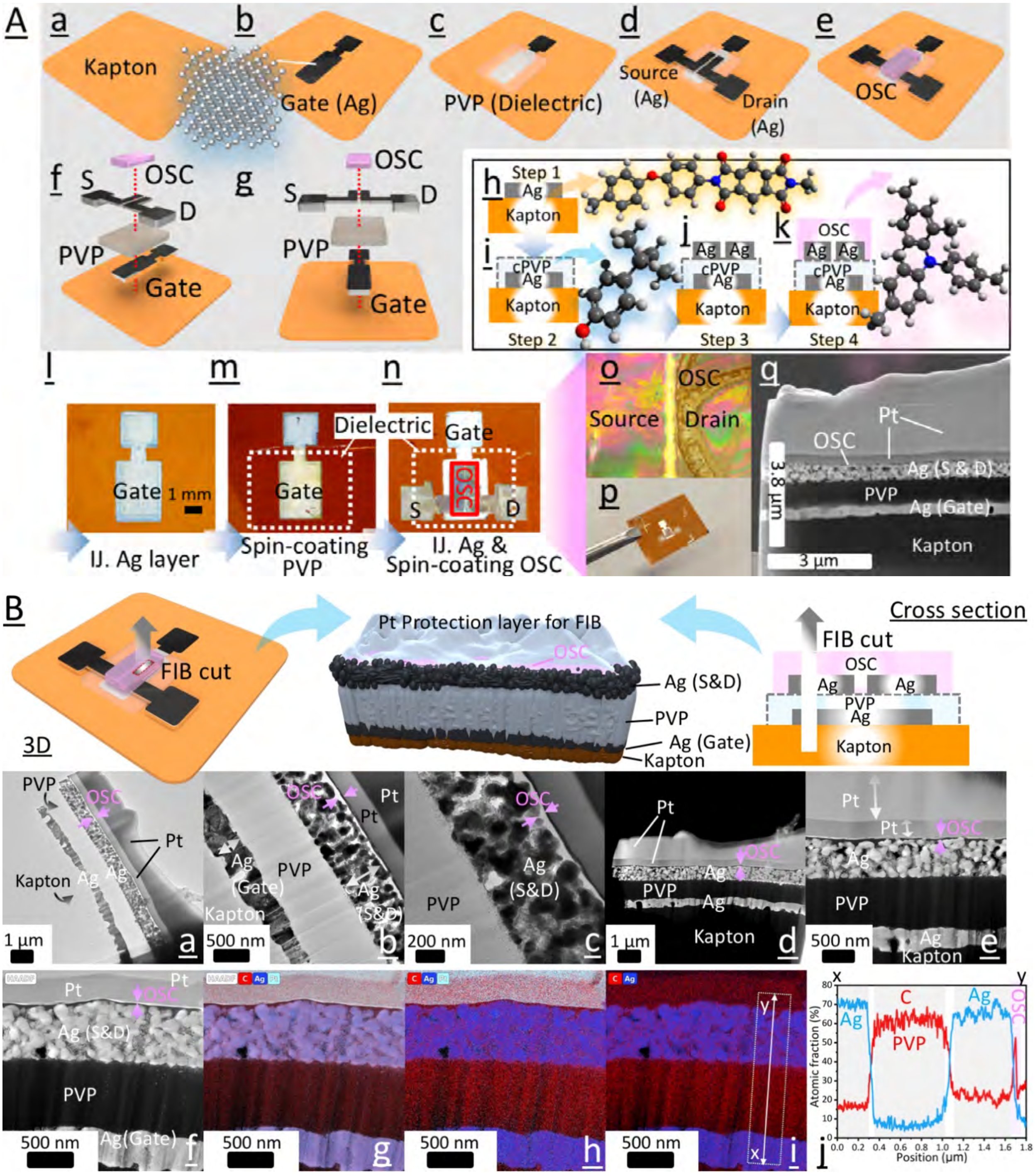

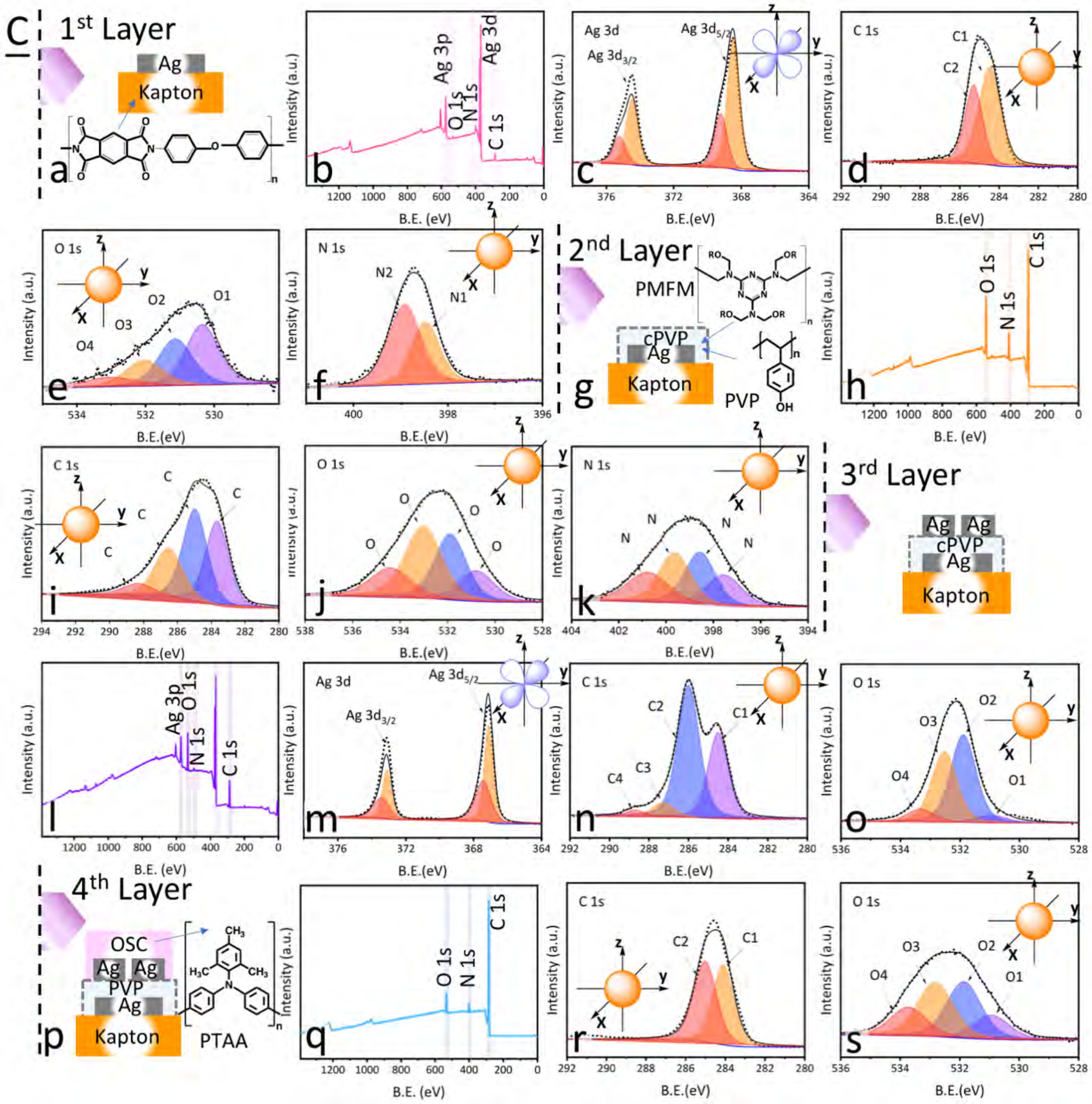

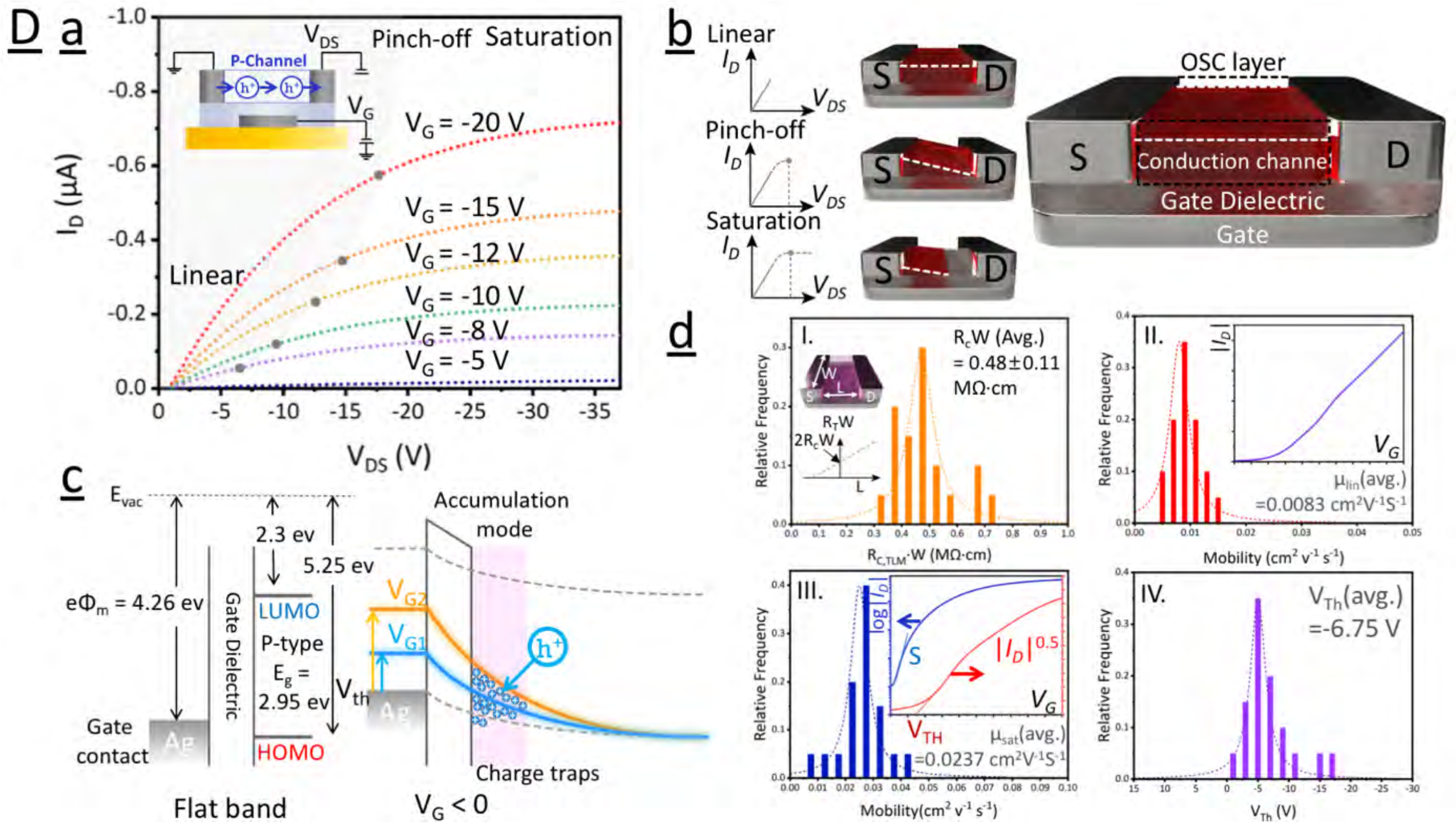

**Figure 7.** A) a-g) Construction of the four different layers in the organic TFT. h-k) Composition and molecular structures of the four layers in the TFT. l-n) Optical images of the constructions of the TFT. o) Magnified images of the OSC layer on top of the source/drain regions in the TFT. p) Appearance of the TFT. q) SEM image of the FIB sample from the TFT. B) a) Low-magnification TEM image for the FIB lift-out TFT sample with a 1 μm scale bar. b) TEM image for the FIB lift-out TFT sample with a 500 nm scale bar. c) TEM image for the FIB lift-out TFT sample with a 200 nm scale bar. d) High-angle annular dark-field (HAADF) STEM image for the FIB cut TFT sample with a 1 μm scale bar. e) HAADF-STEM image for the FIB lift-out TFT sample with a 500 nm scale bar. f) HAADF-STEM image for the FIB lift-out TFT sample from a specific area in (e). g) EDS mapping for C, Ag, and Pt with the original HAADF-STEM image. h) EDS mapping for C, Ag, and Pt. i) EDS mapping for C and Ag. j) EDS line scan from x to y showing elemental variation profiles in (i). C). Step-by-step schematic illustration of the OSC/Ag/cPVP/Ag sample (a, g, l and p) and the collected XPS scan spectra: survey spectrum (b, h, l and q) and the decomposition of the high-resolution spectra for the Ag 3d (c, m), C 1s (d, i, n and r), O 1s (e, j, o and s) and N 1s (f, k and t) core levels. The data were recorded based on a bottom-up analysis strategy starting from the first Ag layer (b-f), cPVP/Ag layer (h-k), Ag/cPVP/Ag (m-o) and ending with the OSC/Ag/cPVP/Ag structure (q-t). D) a) P-type organic TFT output characteristics. The inserted plot shows the operation, build, and connections of a TFT. b) Schematic BGBC structure. The inserted graph shows that the conductance channel varies when the operation of a TFT is in the linear, pinch-off, and saturation regimes. c) Band diagram of the Ag-cPVP-p-

type OSC built in the (left) flat band condition and in the (right) accumulation mode. d) I. Extraction of the normalized contact resistance by the transmission line method (TLM) and the histograms of the variation in contact resistance from 20-TLM sets of TFTs. II. Extraction of the carrier mobility for the linear regime, $\mu_{lin}$, from the transfer characteristics (inserted plot) and the histograms of the variation in $\mu_{lin}$ from 20 optimal TFTs. III-IV. Extraction of the carrier mobility for the saturation regime, $\mu_{sat}$ and threshold voltage, $V_{th}$, from the transfer characteristics (inserted plot) and the histograms of the variation in $\mu_{sat}$ from 20 optimal TFTs.

# Supplementary Materials

L. -K. Lin,* J. -T. Tsai,† S. Díaz-Amaya,† M. R. Oduncu,† Y. Zhang,† P. -Y. Huang,† C. Ostos,† J. P. Schmelzel,† R. Mohammadrahimi, † P. Xu, N. Raghunathan, X. Zhang, A. Wei, D. Bahr, D. Peroulis, L. A. Stanciu*

**From Top to Bottom, Simple and Cost-Effective Methods to Nanopattern and Manufacture Anti-Delaminating, Thermally Stable Platforms on Kapton HN Flexible Films Using Inkjet Printing Technology to Produce Printable Nitrate Sensors, Aptasensors, Protein Sensors, and p-Type Organic Thin Film Transistors**

Table S1. Experimentally determined binding energies of Au4f, Ag3d, C1s, O1s and N1s core levels obtained from the fitted XPS peaks. All binding energies of the PE peaks are given with an accuracy of ±0.1 eV.

| PE core level | SOS (eV) | Peak label | DNA/BSA/Au | | Ag | | cPVP/Ag | | Ag/cPVP/Ag | | OSC/ Ag/cPVP/Ag | |
|---|---|---|---|---|---|---|---|---|---|---|---|---|
| | | | BE (eV) | FHWM | BE (eV) | FHWM | BE (eV) | FHWM | BE (eV) | FHWM | BE (eV) | FHWM |
| Au4f7/2 | 3.67 | Au4f | 83.52 | 0.82 | | | | | | | | |
| | | | | | | | | | | | | |
| Ag3d5/2 | 6.00 | Ag3d | | | 368.47 | 0.85 | | | 367.11 | 0.63 | | |
| | | | | | 369.21 | 0.89 | | | 367.44 | 1.09 | | |
| C1s | | C1 | 284.59 | 1.39 | 284.48 | 1.31 | 283.66 | 1.63 | 284.51 | 1.20 | 284.12 | 1.25 |
| | | C2 | 285.90 | 1.37 | 285.29 | 1.05 | 284.97 | 1.71 | 286.00 | 1.31 | 285.02 | 1.38 |
| | | C3 | 287.20 | 1.47 | | | 286.53 | 1.97 | 287.29 | 1.29 | | |
| | | C4 | 288.18 | 1.45 | | | 288.31 | 2.20 | 288.69 | 1.05 | | |
| O1s | | O1 | 530.49 | 1.24 | 530.35 | 1.15 | 530.79 | 1.77 | 530.95 | 1.25 | 530.94 | 1.44 |
| | | O2 | 531.23 | 1.17 | 531.13 | 1.20 | 531.88 | 1.58 | 531.89 | 1.01 | 531.87 | 1.38 |
| | | O3 | 532.10 | 1.20 | 532.04 | 1.30 | 533.02 | 1.84 | 532.53 | 1.00 | 532.82 | 1.43 |
| | | O4 | 533.00 | 1.30 | 533.04 | 1.72 | 534.41 | 2.02 | 533.33 | 1.05 | 533.72 | 1.40 |
| N1s | | N1 | 397.77 | 1.57 | 398.48 | 0.72 | 397.55 | 1.80 | | | 398.68 | 0.97 |
| | | N2 | 399.04 | 1.23 | 398.92 | 0.82 | 398.57 | 1.55 | | | 399.44 | 0.97 |
| | | N3 | 399.87 | 1.21 | | | 399.62 | 1.61 | | | 400.04 | 0.95 |
| | | N4 | 400.75 | 1.24 | | | 400.78 | 2.04 | | | | |

BE- Binding energy; FHWM – Full width at half maximum; SOS – spin-orbit splitting.

Supplementary Experiment 1: (Printing substrate selection)
Printing substrate plays a significant role in 2D integrated circuit fabrications since it is related to surface property, adhesion, curing temperature, and circuit flexibility. The substrates with Au/Ag printing traces, which need high sintering temperatures, in Table S2 were examined by the whole fabrication process, and their performance was investigated to fabricate a stable sensing platform. First, four kinds of substrates, including Novele paper, silicon wafer with oxide layer, glass slide, and Kapton, were examined with curing process, delamination examination process, and conductivity test for choosing an optimal substrate for matching with Novacentrix 40 wt% silver and UT Dots 20 wt% gold inks. The result showing Kapton HN film had optimal thermal stability and adhesion with both Ag/Au inks and was promising for further investigation. [1]

Table S2 Printing substrate selection [1]
a. NoveleTM + Ag; b. NoveleTM + Au; c. silicon wafer with oxidation layer + Ag; d. silicon wafer with oxidation layer + Au; e. glass slides + Ag; f. glass slides + Au; g. Kapton HN film + Ag; h. Kapton HN film + Au.

| Substrate and ink | Curing at higher than 250 °C | In deionized water | In ethanol | Conductivity |
|---|---|---|---|---|
| a. | damaged | N/A | N/A | N/A |
| b. | damaged | N/A | N/A | N/A |
| c. | good condition | slight delamination happened with fully sintered patterns | slight delamination happened with fully sintered patterns | Yes |
| d. | good condition | serious delamination happened with fully sintered patterns | serious delamination happened with fully sintered patterns | Yes |
| e | good condition | intermediate delamination happened with fully sintered patterns | intermediate delamination happened with fully sintered patterns | Yes |
| f | good condition | serious delamination happened with fully sintered patterns | serious delamination happened with fully sintered patterns | Yes |
| g | good condition | good condition | good condition | Yes |
| h | good condition | good condition | good condition | Yes |

Supplementary Experiment 2: (Inkjet printing ink curing/sintering)
It is known that the curing/sintering temperature plays an important role in inkjet printing pattern fabrication since the curing process directly reflects on the electrical properties. Nanoparticle-based ink has a higher curing temperature, so the curing temperature must be properly adjusted to make printing patterns fully sintered. Incomplete sintering process leads to non-conducting printing patterns, while over-sintering damages the patterns. For the silver ink, the temperatures tested were 190, 230, 280, 330, and 400 °C, and the times tested were 5, 10, 15, 20, and 25 mins. For the gold ink, the temperatures tested were 190, 230, 250, 300, and 400 °C, and the times tested were the same as that of silver. The SEM images used to monitor the curing process are shown in Figure S1 (1) for both temperature control and time control. The growing metal grains in groups A and C with increasing temperature were observed in both metal traces and reached the largest sizes at 400 °C. In Figure S1 (1) A and C, an obvious difference in grain size was observed within d and e samples, which means there should be a critical curing temperature between 330 and 400 °C for the silver ink and between 300 and 400 °C for the gold ink. Therefore, 330 and 300 °C were chosen for the silver and gold inks, respectively, for further investigation into the time. In Figure S1 (1) groups B and D, the growing grains were observed when the curing time was prolonged from 5 to 25 mins. However, the grain sizes for both inks could not reach the sizes when it was at 400 °C, which means the critical temperature should be higher than 330 °C for the silver nano flake ink and 300°C for the gold nanoparticle ink. The curing processes of silver and gold traces were also monitored by XRD in Figure S1 (2).

The electrical properties for each group was examined and cross-verified with SEM images by Four Point Probe Resistivity Measurements in Figure S1 (3). The resistivity values in Figure S1 (3) were calculated with the thickness value obtained from Figure S1 (4) with a profilometer. The profilometry along the x direction for both traces were acquired and the regions highlighted with blue legends were averaged to be the thickness of the films as 786.6 nm for the silver and 477.3 nm for the gold. The obtained resistivity was coherent to grain sizes in the corresponding SEM images, which showed that the grain sizes went up with decreasing resistivity since the metal film became more condensed. After several trials, 400 and 330 °C were chosen to be the optimal curing temperature for the silver and gold, respectively. [1]

Notably, during the profilometry measurements, we noticed the thickness values of metals changed with different printing patterns, so each experiment was calculated separately based on their thickness.

1

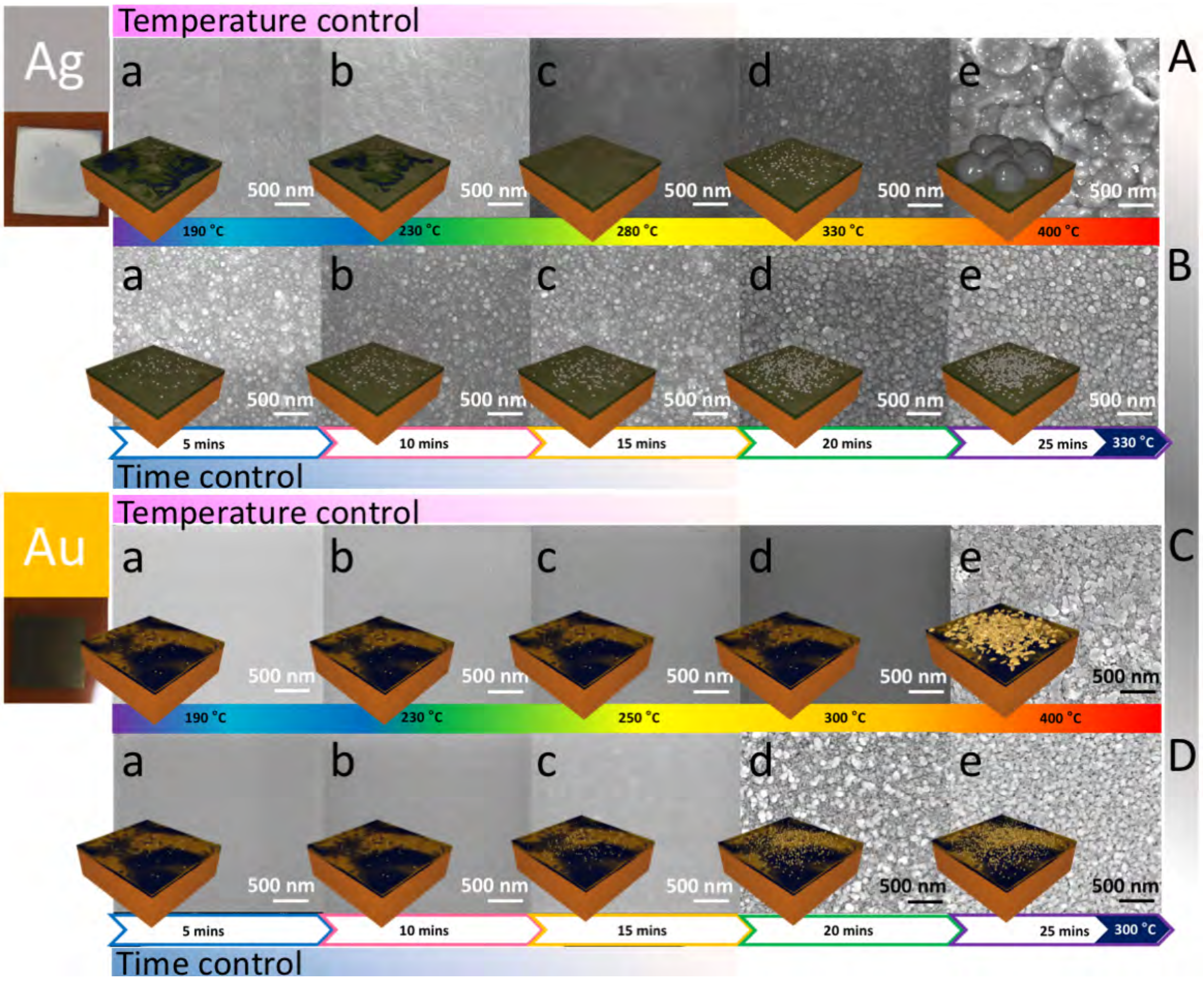

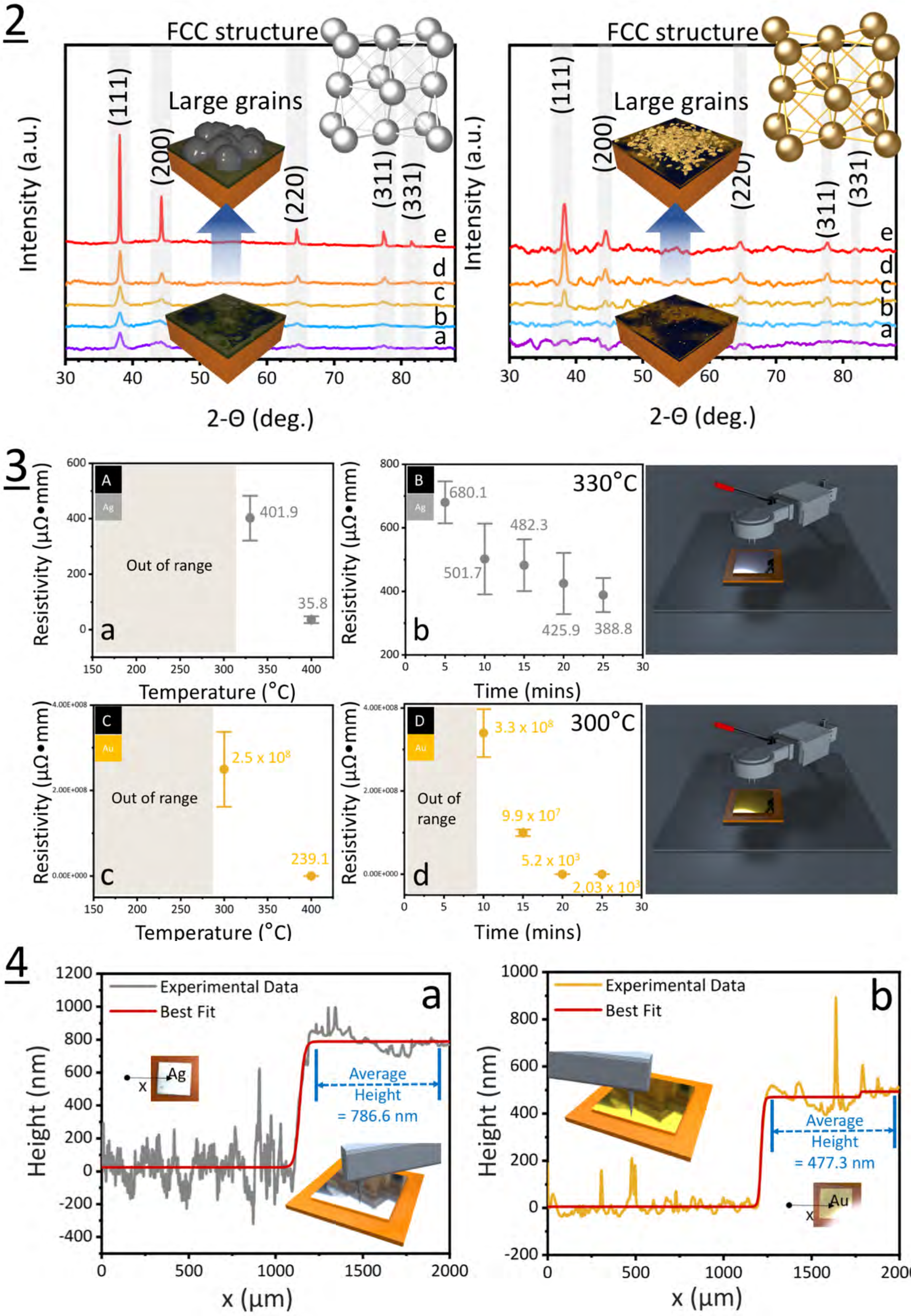

Figure S1 (1) SEM images of traces under inkjet printing curing processes controlled by temperature (A from legends a to e) at 190, 230, 280, 330, and 400 °C for Ag and (C from legends a to e) at 190, 230, 250, 300, and 400 °C for Au, and process controlled by time (B

from legends a to e for silver ink and D from legends a to e for gold ink) within 5, 10, 15, 20, and 25 mins. (2) (left) XRD for printed Ag traces and (right) XRD for printed Au traces in (1). (3) Resistivity of the silver traces against (a) curing temperature and (b) curing time at 330 °C, and of the gold traces against (c) curing temperature and (d) curing time at 300 °C. (4) The profilometry data of (a) silver trace and (b) gold trace scanned along the x direction in the graphs. [1]

Supplementary Experiment 3: (Two probe resistance measurement of inkjet printing traces)
The linear resistors were used to estimate resistance against trace width to predict the performance of a linear circuit (Figure S2). Fixed parameters were used for printing the traces, and the resulting traces for silver from 60 to 550 μm and for gold from 30 to 450 μm were measured. The short circuits for gold traces were found when the width was smaller than 40 μm. The silver traces basically had resistance below 60 ohms due to the higher metal percentage in the ink. The further sensing fabrication for gold traces were then confined to over 50 μm to avoid short circuits. [1]

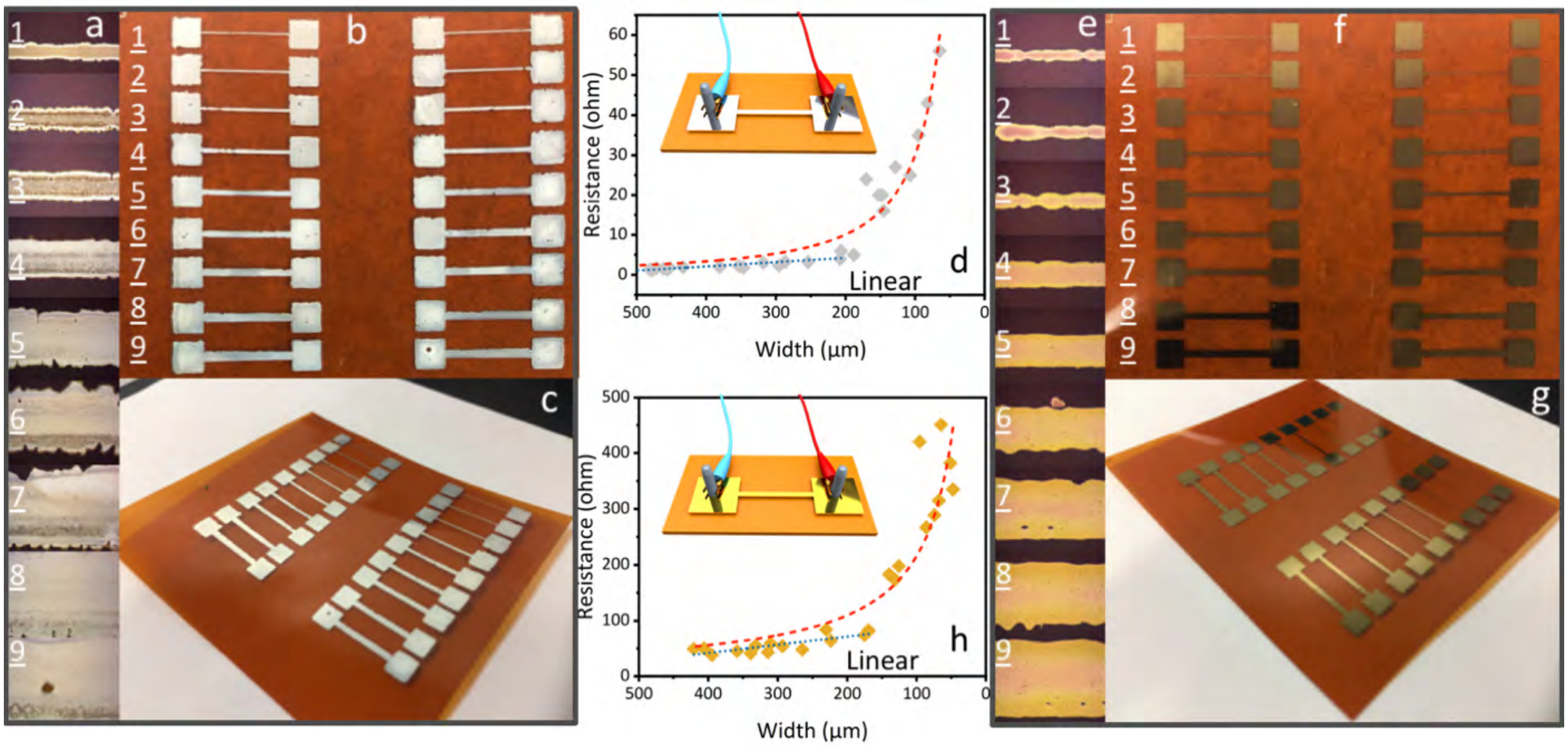

Figure S2 Resistance of linear resistors determined by a two-probe measurement. The images from a 10x optical microscope are of (a) silver resistors with (1) 80.7, (2) 127.7, (3) 156.2, (4) 201.1, (5) 286.2, (6) 342.9, (7) 410.6, (8) 463.1, and (9) 494.9 μm in average width, and (b) gold resistors with (1) 50.5, (2) 66.1, (3) 69.4, (4) 132.4, (5) 171.4, (6) 239.7, (7) 311.9, (8) 337.3, and (9) 407.3 μm in average width; photographs of (b) (c) silver resistors and (f) (g) gold resistors; The two-probe resistance values against the (d) silver trace width and (h) gold trace width. [1]

Supplementary Experiment 4: (Thin film mechanical properties measurements and simulations)
The foregoing sections compare the experimentally measured and calculated normalized resistance following the inspection line from the coated coupon, see Figure S3 A. The location of inspection lines for normalized resistance are chosen to match the location of the experimental results.

For the finite element model, boundary conditions appropriate for modeling of the tensile test and four-point bending test were applied to the virtual coupon. In the physical experiment, a coupon is gripped within the MTS machine. To better reflect the Poisson's effects during the tension test simulation, the gripped regions were also modeled, and the corresponding boundary conditions were applied on them. For the simulation, the bottom "gripped" area was constrained against all displacements and rotations, while the upper "gripped" area was uniformly pulled with a prescribed displacement $u^*$ thus causing the specimen to develop a reaction force $F_x(u^*)$. The reference points were created for convenience of applying the boundary conditions and reading of the analysis results. The degrees of freedom of "gripped" regions were tied to the degrees of freedom of the reference points. The coated film macroscopic (global/far-field) stress ($\bar{\sigma}_{xx}$) and strain ($\bar{\epsilon}_{xx}$) are calculated as given by Eq. (1):

$$\bar{\epsilon}_{xx} = \frac{u^*}{L}; \; \bar{\sigma}_{xx} = \frac{F_x}{wt} \tag{1}$$

where L, w, t are the coupon gage length, width and thickness, respectively.

The volume resistivity (electrical resistivity) was measured by a four-probe apparatus. This method can measure the volume resistivity of a thin film or coated line on a substrate. The four-probe was designed with the outer two probes measuring current and the inner two probes measuring voltage. Several coated distances were measured to determine the volume resistivity per-cross-sectional area by utilizing the following equation (2):

$$R = \rho \frac{\ell}{A} \tag{2}$$

Where R is the electrical resistance of the coat layer on the substrate calculated from the measured current and voltage; ρ is the electrical volume resistivity; $\ell$ is the length of the specimen; A is the cross-section area of the silver or the gold film.

Supplementary Note 5: Mechanical Characterization of Thin Films (Nanoindentation Experiment):

Nanoindentation experiment was carried out using Hysitron TI950 (Hysitron Inc., Minneapolis, MN) fitted with Berkovich probe with the effective radius of 300 nm. The probe shape is schematically shown in the supplementary. Bekovich probe is a three-sided pyramid probe which is geometrically self-similar. The quasi-static single load function with maximum load of 18 uN was defined and applied on both samples. The hardness of thin films was directly calculated by the Oliver–Pharr method using the following equation (3): [2]

$$H = \frac{P_{max}}{A_c} \tag{3}$$

Where $P_{max}$ is the maximum applied load and $A_c$ is the projected contact area which is a function of contact depth and the probe shape.

The nanoindentation results are shown in Figure S3 B. Figure S3 B a demonstrates the typical load-displacement curve which is obtained after applying a complete cycle of loading-holding-unloading on a specimen. The hardness plot versus the contact depth ($h_c$) of both indented thin films are shown in Figure S3 B b and B c. It is clearly observed that by applying the same maximum load on Ag and Au thin films, the average contact depth is shallower in Ag thin film in compared with Au thin film. As a result, the hardness values of Ag thin film and Au thin film in the Figure S3 B e and B f are calculated as 1.784 GPa and 0.733 GPa. Therefore, according to the nanoindentation results, Au thin film is softer than Ag thin film. [2]

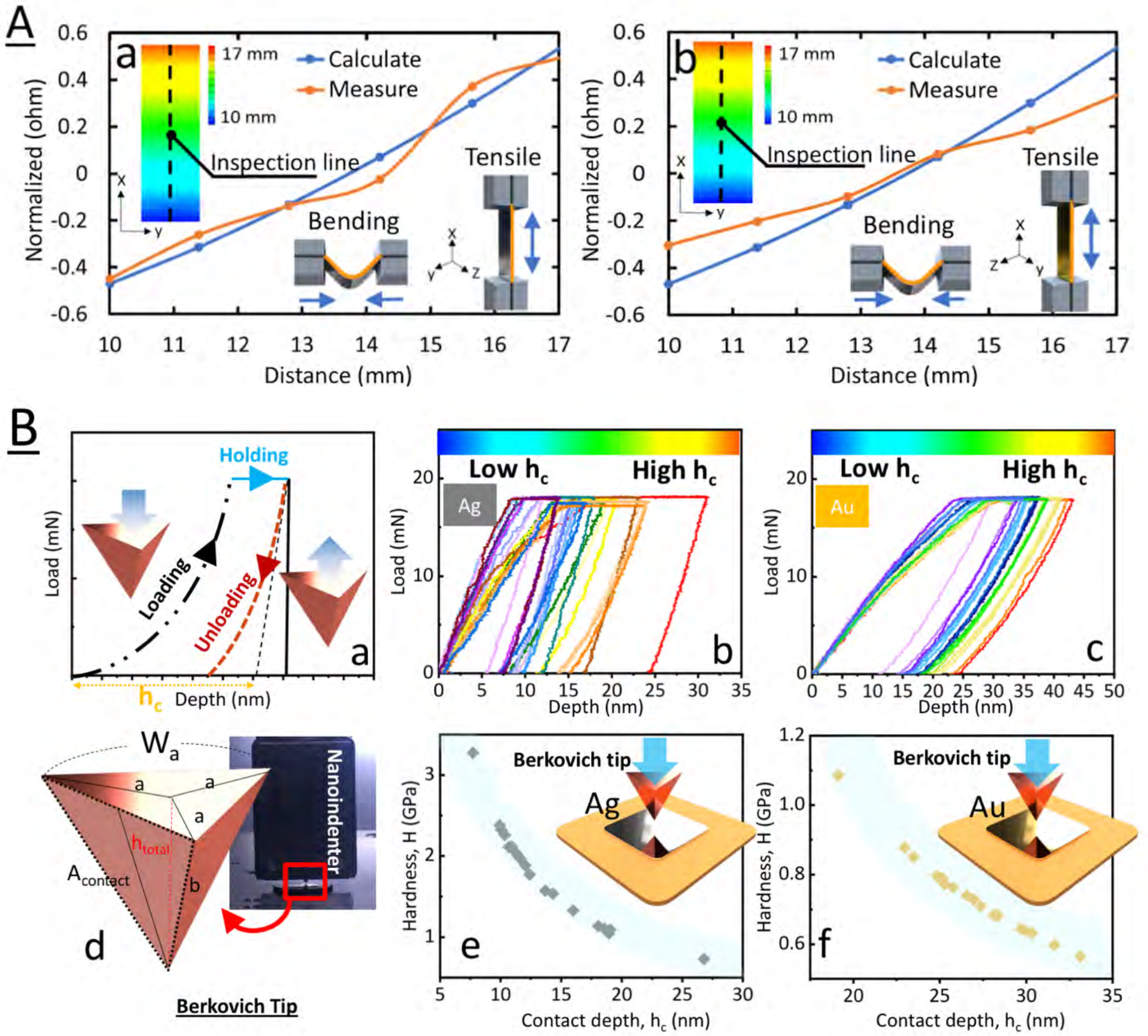

Figure S3 (A) Normalized resistance under mechanical testing. (a) Mechanical test results of Ag of film. (b) Mechanical test results of Au film. (B) nanomechanical property analyses: (a) Schematic typical load-displacement curve of an elasto-plastic material. (b) Load-displacement curves resulted from indenting Ag thin film. (c) Load-displacement curves resulted from indenting Au thin film. (d) Schematic shape of Berkovich probe. (e) Hardness vs contact depth plot which is resulted from indenting Ag thin film. (f) Hardness vs contact depth plot which is resulted from indenting Ag thin film.

Supplementary Experiment 6: (analytical performance for electrochemical $Hg^{2+}$ sensors) Standard solutions of mercury were purchased from Sigma Aldrich. DI water were used to dilute mercury solution into 0 to 100 ppm stock solutions with known concentrations. A drop of 10 µl $Hg^{2+}$ solution was acquired from the stock solutions and incubated with the surface of the printed working electrodes for 15 mins at 25 °C, and the electrodes were wash with DI water. During the process, DI water was used to replace mercury solution to control measurement background. PEIS was adopted for characterize the Printed $Hg^{2+}$ sensor.

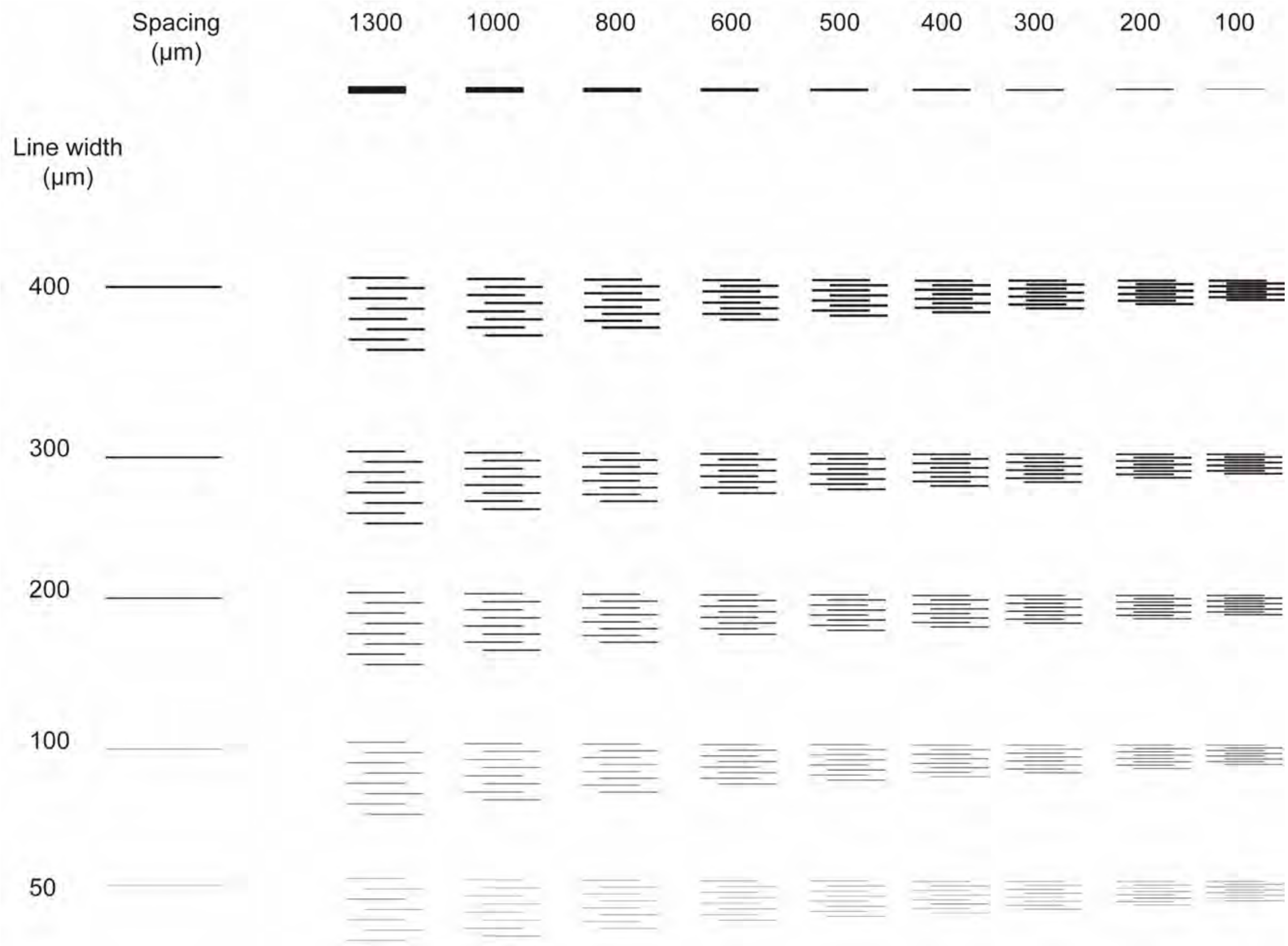

Figure S4. Printed trace optimization sheet

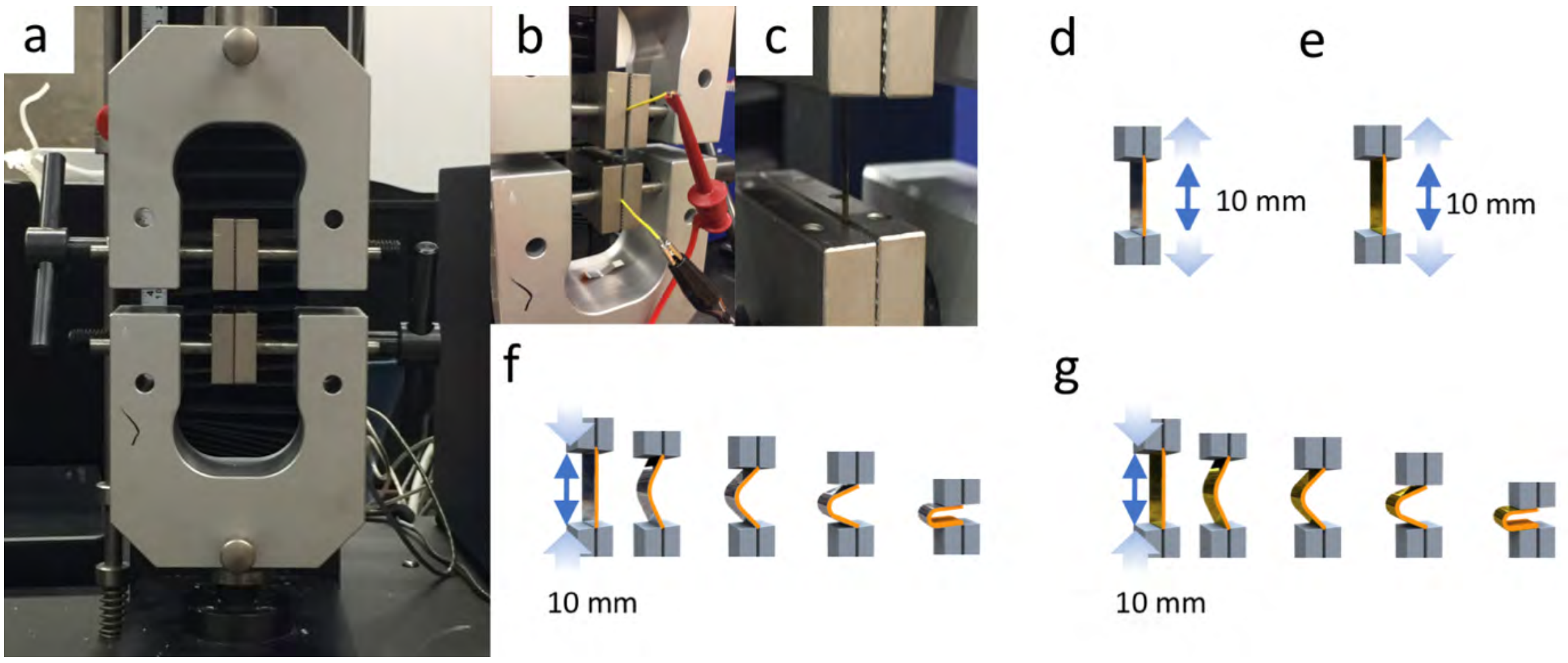

Figure S5. Scheme of thin film conductivity tests thought stretching and bending

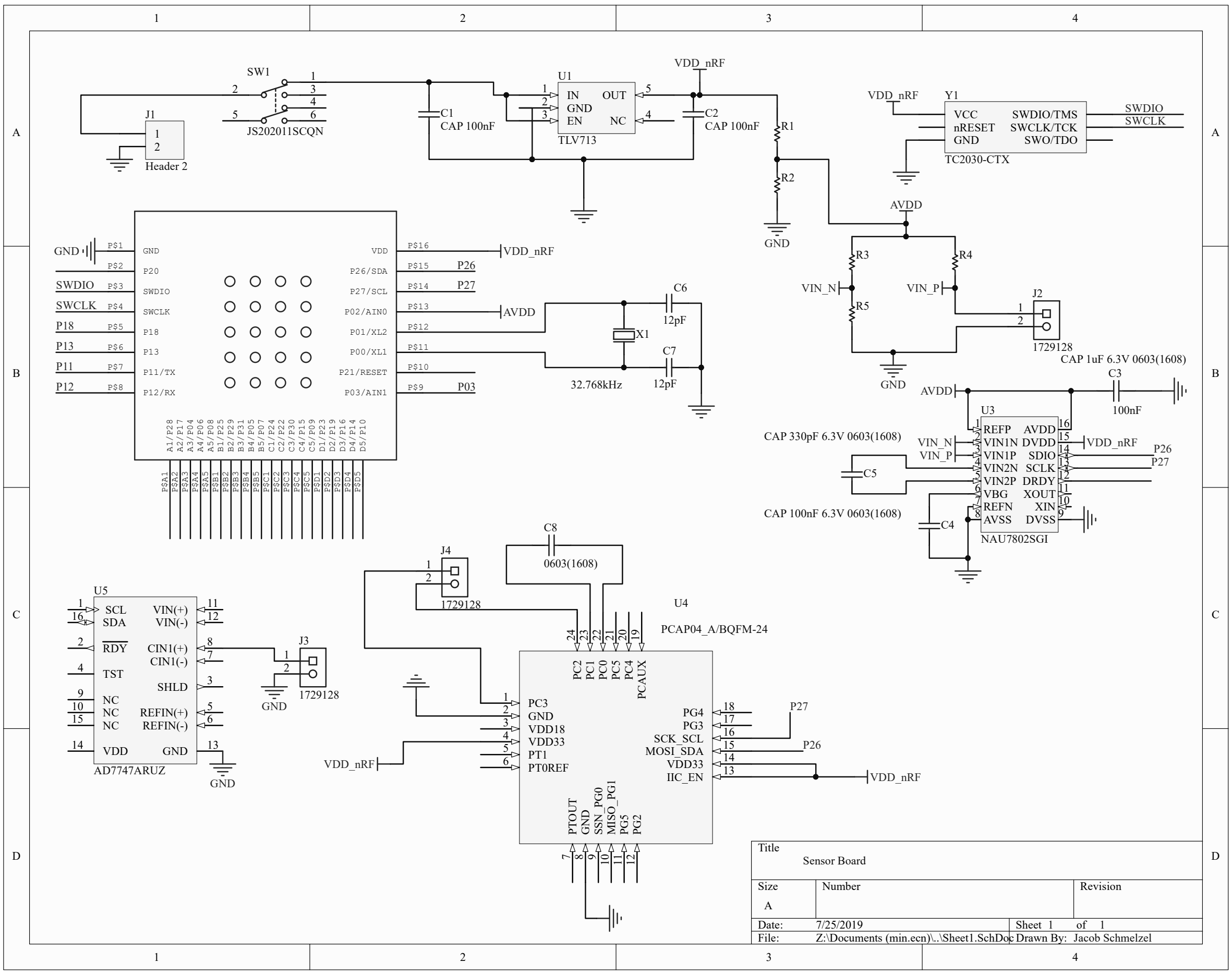

Figure S6. Blueprint for the sensor PCB (designed by Jacob Peter Schmelzel)